\documentclass[11 pt]{article}
\usepackage[margin=1in]{geometry}
\usepackage{graphicx}
\usepackage{amssymb}
\usepackage{amsmath}
\usepackage{amsthm}
\usepackage{mathrsfs}

\usepackage[usenames,svgnames]{xcolor}
\usepackage{multirow}
\usepackage{rotating}
\usepackage{colortbl}

\usepackage[numbers,square,comma,sort&compress]{natbib}
\usepackage[normalem]{ulem}
\usepackage[colorlinks]{hyperref} 
\usepackage{lineno}
\usepackage[colorinlistoftodos,backgroundcolor=LightGrey,linecolor=DimGrey]{todonotes}

\usepackage{tikz-cd}

\theoremstyle{plain}
\newtheorem{theorem}{Theorem}[section]
\newtheorem{lemma}{Lemma}[section]
\newtheorem{proposition}{Proposition}[section]

\theoremstyle{definition}
\newtheorem{definition}{Definition}[section]
\theoremstyle{remark}
\newtheorem{remark}{Remark}[section]
\newtheorem{example}{Example}[section]

\providecommand{\R}{\mathbb{R}}

\providecommand{\C}{\mathbb{C}}

\providecommand{\chain}{\mathscr{C}}
\providecommand{\set}{\mathscr{S}}

\providecommand{\mb}[1]{\mathbf{#1}}

\newcommand{\beginsupplement}{%
        \setcounter{section}{0}
        \renewcommand{\thesection}{A\arabic{section}}%
        \setcounter{table}{0}
        \renewcommand{\thetable}{A\arabic{table}}%
        \setcounter{figure}{0}
        \renewcommand{\thefigure}{A\arabic{figure}}%
        \pagebreak
        \noindent{\Large\textbf{Appendix}}
     }

\newcommand{\captionfonts}{\small}
\makeatletter  
\long\def\@makecaption#1#2{%
  \vskip\abovecaptionskip
  \sbox\@tempboxa{{\captionfonts #1: #2}}%
  \ifdim \wd\@tempboxa >\hsize
    {\captionfonts #1: #2\par}
  \else
    \hbox to\hsize{\hfil\box\@tempboxa\hfil}%
  \fi
  \vskip\belowcaptionskip}
\makeatother   

\newcommand{\inlinesec}[1]{\medskip\noindent\textbf{#1.}}

\newcommand{\comment}[1]{}

\newcommand{\missingcite}[1]{\textcolor{red}{[cite]}}

\title{\sffamily Input-output equivalence and identifiability: some simple generalizations of the differential algebra approach}
\author{Marisa C. Eisenberg\footnote{Departments of Complex Systems, Epidemiology, and Mathematics, University of Michigan, Ann Arbor}}

\date{}

\begin{document}


\maketitle

\begin{abstract} \sffamily
In this paper, we give an overview of the differential algebra approach to identifiability, and then note a very simple observation about input-output equivalence and identifiability, that describes the identifiability equivalence between input-output equivalent models. We then give several simple consequences of this observation that can be useful in showing identifiability, including examining non-first order ODE models, nondimensionalization and rescaling, model reducibility, and a modular approach to evaluating identifiability. 
We also examine how input-output equivalence can allow us to generate input output equations in the differential algebra approach through a wider range of methods (e.g. substitution and differential or standard Groebner basis approaches).\footnote{This paper still in draft form---currently it serves mostly as a place to keep several simple but useful results that come up frequently in proving identifiability, and to provide an introduction to the differential algebra method.}
\end{abstract}

\section{Introduction}
Identifiability analysis addresses the question of whether it is possible to uniquely recover the parameters from a given set of data.  This problem can be broken into two broad (and sometimes overlapping) categories---practical or numerical identifiability, which incorporates practical estimation issues such as sampling times, noise, and bias, and structural identifiability, which considers a best-case scenario when the data are assumed to be known completely (i.e. smooth, noise-free, and known for every time point in the case of differential equation models).  Structural identifiability is a necessary condition for parameter estimation with real data, and can yield information about how to reparameterize the model when it is unidentifiable.  

Many different approaches to structural identifiability analysis have been developed \cite{Bellu2007, Chappell1998, Cobelli1980, Pohjanpalo1978}.  However, the computational intensity of many methods makes applications beyond relatively simple models challenging \cite{Saccomani2003}.  For linear models, identifiability can be determined globally via a transfer function approach and other linear algebra methods \cite{Bellman1970, Cobelli1980, DiStefano1983}.  One successful approach to identifiability for polynomial and rational function ordinary differential equation (ODE) models is via differential algebra \cite{Bellu2007, Ljung1994, Saccomani2003}, which can be used to determine not only the overall identifiability of the model, as well as the identifiable parameter combinations in the case of model unidentifiability, which can be used to find reparameterizations of the model in terms of these combinations \cite{Meshkat2009}.   

The usual algebraic approach \cite{Ljung1994,Ollivier1990,Audoly2001,Saccomani2003} is based on using characteristic sets \cite{Ritt1950} (a method of solving/reducing differential polynomial systems) to generate a monic set of simplified equations in terms of only the known or measured variables and the parameters, called the \emph{input-output equations}, whose solutions are the set of all input-output pairs for the model.  The coefficients of the input-output equations can then be used to test identifiability of the model \cite{Audoly2001,Saccomani2003,Margaria2001}.  One of the major limitations of this approach is that for more complex models it can become computationally intractable \cite{Saccomani2003}. Recently a faster approach has been developed that uses the differential algebra method to evaluate identifiability without generating input-output equations \cite{hong2018global}. Nonetheless, input-output equations remain useful, particularly for proving results for identifiability of general classes of models. 
Thus, expanding the approaches to generate input-output remains an important goal. Intuitively one would expect that what makes the input-output equations informative about model identifiability is not that they are generated by characteristic sets per se, but rather that their solutions are the solution trajectories of the measured variables of the original system. Moreover, there are many different approaches to reducing or solving systems of equations (e.g. Gr\"{o}bner bases, \textit{ad hoc} substitution), which intuitively ought to generate equivalent results (and indeed Meshkat et al. \cite{Meshkat2012} showed that Gr\"{o}bner bases can be used to generate input output equations). It would therefore seem natural to extend this approach to more general differential algebraic methods of generating input-output equations or evaluating identifiability. 

A very closely related property to identifiability is that of \emph{distinguishability} or input-output equivalence, which addresses the question of whether two models generate the same set of input-output pairs (if so, they are termed indistinguishable). An unidentifiable model (for example) is indistinguishable from its identifiable reparameterization.
Distinguishability is in some sense similar to the identifiability question, but examines whether there are multiple ways to generate a given output at the overall model scale. 
Because identifiability depends on whether multiple parameter sets can generate the same input-output pairs, we would expect that indistinguishable models should have related identifiability properties. 

In this paper, we give an overview of the differential algebra approach to identifiability, and then note a very simple observation that describes the identifiability equivalence between input-output equivalent models. We then note several simple consequences of this observation that can be useful in showing identifiability, including examining higher order ODE models, nondimensionalization and rescaling, model reducibility, and a modular approach to evaluating identifiability. 
Finally, we examine how input-output equivalence can allow us to generate input output equations in the differential algebra approach through a wider range of approaches, including simple substitution and differential or standard Groebner basis computations \cite{Mansfield1991, Meshkat2012}.

\section{Identifiability}
We begin by introducing the identifiability framework used here. Let the model be given in state-space form \cite{Audoly2001, Ljung1994} by:
\begin{equation}
\begin{aligned}
\dot{\mb{x}} &= f(\mb{x},t,\mb{u,p})\\ 
\mb{y} &= g(\mb{x},t,\mb{p})
\end{aligned}
\label{eq:modelsetup}
\end{equation}
where $\dot{\mb{x}}$ is a system of first order ordinary differential equations (ODEs), with $t$ representing time, and $\mb u$ the experimental input function(s), if any.  The model parameters are given by the $n_p$ dimensional vector $\mb{p} \in \R^{n_p}$ (the complex numbers $\C$ may also sometimes be considered, depending on the model). We will occasionally refer to individual parameters within $\mb p$ as $p$ (without bold-face), and refer to an arbitrary point in parameter space ($\R^{n_p}$) as $\mb p^*$. The measured data/output(s) are given by $\mb{y}$, which represents the the $n_y$-dimensional vector of output variables without any measurement error.  We also let $\mb{x}(0)$ represent the vector of initial conditions for $\mb x (t)$.  In some cases, the initial conditions (or some of them) will also be included as unknown parameters in $\bf p$. 

As in \cite{Audoly2001,Bellu2007}, we assume that $f$ and $g$ are rational polynomial functions of their arguments, and that $\bf u,x$, and $\bf y$ are arbitrarily differentiable.  We also assume that any constraints reflecting known relationships among parameters, variables, inputs, and outputs 
are either already included in the model equations or are appended to them, as these are known to affect identifiability properties \cite{Meshkat2009,Cobelli1980,DiStefano1983}. In this paper, we will examine only equality constraints, although inequality constraints are also sometimes examined \cite{meshkat2014finding}.

We will refer to the collection of state equations (in this case ODEs), measurement equations, and any constraint equations (i.e. Eq.~\eqref{eq:modelsetup} as well as any constraints), as a \textit{model}. 
We will denote a model by $M(t,\bf u,x,y,p)$ or sometimes $M(\bf p )$ or simply $M$ for shorthand. We also note that while we write Eq.~\eqref{eq:modelsetup} as first order ODEs here, ODE models of higher order can also be considered by rewriting them as a system of first order equations (discussed further below). 

Structural identifiability analysis in this context explores the question: given a model, 
is it possible to uniquely identify the parameters $\mb{p}$, assuming ``perfect'' noiseless data?  Mathematically, this can be thought of in terms of injectivity of the \emph{model map} $\Phi: \bf p \mapsto y$ given by viewing the model output $\mb{y}$ as a function of the parameters $\mb{p}$ and the known inputs $\mb{u}$ (if any) \cite{Meshkat2009, Saccomani2003}.  We note that because there may be some `special' or degenerate parameter values or initial conditions for which an otherwise identifiable model is unidentifiable (e.g. if all initial conditions are zero and there is no input to the model), structural identifiability is often defined for almost all parameter values and initial conditions \cite{Meshkat2009, Audoly2001, Saccomani2003}. 

\begin{definition} For a given model with state equations $\dot{\mb{x}} = f(\mb{x},t,\mb{u,p})$ and output  $\mb{y}$, an individual parameter $p$ is \emph{uniquely (or globally) structurally identifiable} if for almost every value $\mb{p}^*$ and almost all initial conditions, the equation $\mb{y}(\mb{x},t,\mb{p}^*) = \mb{y}(\mb{x},t,\mb{p})$ implies $p = p^*$.  A parameter $p$ is said to be \emph{non-uniquely (or locally) structurally identifiable} if for almost any $\mb{p}^*$ and almost all initial conditions, the equation $\mb{y}(\mb{x},t,\mb{p}^*) = \mb{y}(\mb{x},t,\mb{p})$ implies that $p$ has a finite number of solutions.  \end{definition}

\begin{definition}  Similarly, a model with state equations $\dot{\mb{x}} = f(\mb{x},t,\mb{u,p})$ and output  $\mb{y}$ is said to be \emph{uniquely} (respectively \emph{non-uniquely}) \emph{structurally identifiable}  for a given choice of output $\mb{y}$ if every parameter is uniquely (respectively non-uniquely) structurally identifiable, i.e. the equation $\mb{y}(\mb{x},t,\mb{p}^*) = \mb{y}(\mb{x},t,\mb{p})$ has only one solution, $\bf p = p^*$ (respectively finitely many solutions).  Equivalently, a model is uniquely structurally identifiable for a given output if and only if the map $\Phi$ is injective almost everywhere, i.e. if there exists a unique set of parameter values $\mb{p}^*$ which yields a given trajectory $\mb{y}(\mb{x},t,\mb{p}^*)$ almost everywhere. \end{definition}

The equivalence classes generated by $\Phi$ are precisely the sets of parameter values yielding the same output, so that if the fibers of $\Phi$ contain finitely many elements, the model is locally (non-uniquely) identifiable, and if the fibers of $\Phi$ contain infinitely many elements, the model is termed \emph{unidentifiable}. In this case, typically there exists a set of identifiable combinations of parameters that represents the parametric information available in the data (except in degenerate cases where the model is reducible or has insensitive parameters). These combinations are not unique---sets of identifiable combinations that generate the same field are equivalent, e.g., $\{ab, c/b\}$ and $\{ab, ac\}$.

We note that a structurally identifiable model may still be practically unidentifiable for a variety of reasons--- for example, if the model identifiability is highly sensitive to measurement error in the data (denoted \emph{numerical or practical unidentfiability} \cite{Jacquez1985}), or if an incorrect model is used, so that the model structure cannot fit the data (i.e. the measured data is far from the image of $\Phi$ so that the model can never realize the given data).  The first case is often due to output insensitivity in the parameters, wherein small changes in the parameters yield different output trajectories, but only slightly different, so that even very small measurement error can render the model practically unidentifiable.  

\subsection{The differential approach to identifiability using characteristic sets}\label{sec:charset}
Next we provide a brief overview of the differential algebra approach to identifiability using characteristic sets; an overview of some of the fundamentals of differential algebra as related to identifiability is given in the Appendix, and for a full treatment of the differential algebra approach to identifiability we refer the reader to \cite{Ollivier1990, Ljung1994, Audoly2001, Saccomani2003}.

In the usual differential algebra approach to evaluating identifiability \cite{Audoly2001,Ollivier1990, Ljung1994}, one begins by viewing the model equations as differential polynomials in the differential ring $\mb{R}\{t, \bf x, u, y\}$ (where $\bf R$ is the ring of coefficients, including the model parameters), by rewriting them with all terms on one side, as  $\dot{\mb{x}} - f(\mb{x},t,\mb{u,p}), \mb{y} - g(\mb{x},t,\mb{p})$ (clearing denominators as needed). We treat the model equation polynomials as generators of a differential ideal in $\bf R\{x\}$, and then take the characteristic set of this differential ideal \cite{Ritt1950, Ljung1994}. This relies in part on the differential ideal for the model being prime \cite{Ljung1994}, which is ensured if the model is in state-space form (i.e. first order equations of the form in Eq.~\eqref{eq:modelsetup}) \cite{Ljung1994, Audoly2001}.

The characteristic set gives an autoreduced form of the model equations that includes the \emph{input-output equations}, a set of $n_y$ monic differential polynomials in terms of only the observed or known variables ($\bf y$ and $\bf u$), their derivatives, and the model parameters. The input-output equations are an implicit form of the model map $\Phi$ \cite{Audoly2001}, meaning that for a given set of parameters $\mb p$ and inputs $\mb u$, they generate the same output $\mb y$ as the original model \cite{Audoly2001}. We will denote the input-output equations of a model $M$ by $\mb{\Psi}(M)$ or simply $\mb{\Psi}$. 
The coefficients $\bf c(p)$ of the input-output equations form an alternative parameterization of the model, so that we have a commutative diagram:
\begin{center}
\begin{tikzcd}
\mb{p} \arrow[rr, "\Phi"] \arrow[rd] &                    & \mb{y} \\
                                   & \mb{c(p)} \arrow[ru] &  
\end{tikzcd}
\end{center}
where $\bf p\mapsto c(p)$ is the map from the parameters to the coefficients, and $\bf c(p)\mapsto y$ is the map from the coefficient values to an output trajectory, generated by solving the input-output equations for the given coefficient values (i.e. treating the input output equations as differential equations). 

An important piece of the characteristic set approach is that (under relatively mild assumptions of solvability \cite{Saccomani2003}) the coefficients $\bf c(p)$ of the resulting input-output equations are identifiable (i.e. they are identifiable combinations) and contain all the identifiability information in the original model \cite{Audoly2001,Ollivier1990, Ljung1994}. More specifically, the differential algebra approach tells us that if we want to evaluate the injectivity of the model map $\Phi$, we can do so by evaluating the \emph{coefficient map} $\bf p \mapsto c(p)$. We note that this feature of the input-output equations depends on having sufficiently many independent values for $\bf y$ and $\bf u$, so that the coefficients $\bf c(p)$ can be solved uniquely for (i.e. for a given set of values for $\bf y$, $\bf u$, and their derivatives, the input output equations form a linear system in terms of the coefficients, so we need sufficiently many such equations to solve for the coefficients). This condition is termed solvability \cite{Saccomani2003}. In most practical cases, solvability is easily achieved (since we assume we have perfect measurements of $\bf y$ and $\bf u$, providing us with as many points as we like). However, when the dynamics of the model are constrained, the solvability assumption may not hold (e.g. if one or more state variables is actually a constant rather than time varying, as in the example of \cite{hong2018global}). Issues of solvability are rare in practice, but tend to arise when there are constraints that restrict the dynamics to a lower dimensional subspace, but which have not been incorporated explicitly into the model equations. Solvability is typically assumed in the differential algebra approach (in some sense it related to the assumption of `perfect data' used in structural identifiability in general) \cite{Saccomani2003}, and we assume it here.


\section{Identifiability and model input-output equivalence}

Next, we note a simple but useful observation regarding input-output equivalence and identifiability, and from this observation derive several `rules of thumb' which may be useful in establishing identifiability for models. 
In particular, this observation allows us to make some generalizations to the characteristic set approach for evaluating identifiability, by broadening the range of approaches possible to use for generating input-output equations. 

First, we will need the notion of \emph{input-output equivalence}. Input-output equivalence is a condition which ensures that from the perspective of the output variable, the two models (or two formulations of the same model) are dynamically equivalent---in other words, the internal, unobserved variables are in some sense free be altered as we wish, so long as the resulting measured trajectories $\bf y$ remain unchanged. 

Often this is defined by saying that two models are input-output equivalent (also termed \emph{indistinguishable}) if for for any particular input-output pair generated by one model, there exists at least one parameter set for the other model that generates the same input-output pair \cite{raksanyi1985identifiability, walter1996identifiability, distefano2015dynamic}. We note that this definition does not specifically address what initial conditions are needed to generate the desired input-output pair (in part because much of the early literature was based on transfer functions, in settings where the initial conditions were often assumed to be zero), and that this definition does not require us to consider the specific transformation of the parameters from one model to the other. 

Here, we will use the term indistingiuishable to refer to the above idea, and define input-output equivalent in a slightly more specific way, where we explicitly consider the parameter values. Additionally, because there may be some specific points where the two models do not coincide (e.g. if certain parameters or initial conditions are zero), we take our definition to be generic (similar to our definition for identifiability). Let the set of output trajectories for a model given specified parameter values and inputs, but across all allowable initial conditions to be written as $\set(\mb{y})$. Then we define:
\begin{definition}
	Two models $M_1(t, \bf u_1, x_1, y_1, p_1)$ and $M_2(t, \bf u_2, x_2, y_2, p_2)$ 
	are said to be \emph{input-output equivalent} if, given the same parameters and input functions, both models generate the same set of observed trajectories (for almost all initial conditions and parameter values). In other words, $M_1$ and $M_2$ are input-output equivalent if $\bf u_1 = u_2, p_1 = p_2$ implies $\set(\mb{y_1}) = \set(\mb{y_2})$. We write this as $M_1 \sim M_2$.
\end{definition}

This definition aligns with the notions of the model map and input-output relation \cite{Audoly2001}, in that we don't specify any particular form for the internal variables $\bf x$ (or their initial conditions) for either model, so long as the overall mapping from parameters and inputs to outputs is maintained. In most cases, we can explicitly define a conversion of the initial conditions between $M_1$ and $M_2$, allowing us to say these two models are input-output equivalent if for the same $\bf u$, $\mb{x}(0)$, and $\mb{p}$, they generate the same $\bf y$. 

To illustrate the idea of input-output equivalence, we present two examples. 
\begin{example}\label{ex:rescale}
Consider the following two models:
\begin{align*}
& \textrm{Model 1} 				& & \textrm{Model 2}  \\
\dot{x}_1 &= - x_1 + p_1 u(t) 		& \dot{x}_2 &= -x_2 + q u(t) \\
y &= p_2 x_1 				& y &= x_2
\end{align*}
Model 2 can be obtained from Model 1 by rescaling $x$ with $p_2$ (i.e. defining $x_2 = y = p_2 x_1$ and $q = p_1 p_2$, or equivalently by substituting $x_1 = y/p_2$ into the first equation of Model 1 and defining $x_2 = y$). Indeed, Model 2 is just the input-output equation for Model 1, rewritten in state-space form.
To show that they are input-output equivalent, we will rewrite Model 2 so that we can set their parameters and initial conditions to the same values. For this, we append the following equations to Model 2: $q = p_1 p_2$ and $x_2(0) = p_2 x_1(0)$.
Now Model 2 is parameterized in the same way as Model 1, and we see that for all parameters and initial values of $x_1$, Model 2 gives the same output as Model 1, making the two models input-output equivalent. 
\end{example}

This example illustrates how an identifiable reparameterization of a model \cite{Meshkat2012, distefano2015dynamic} represents one example of an input-output equivalent model. Next let us examine a classic example of indistinguishability \cite{distefano2015dynamic}.

\begin{figure}
\centering
\includegraphics[width=0.4\textwidth]{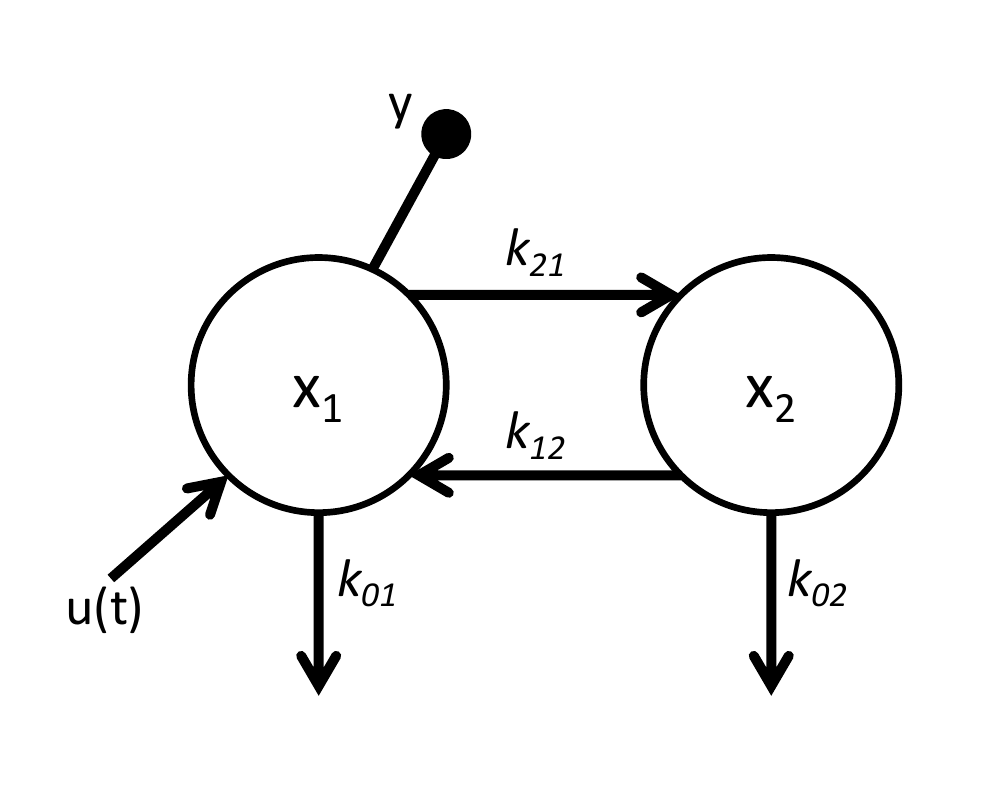}
\caption{Linear 2-compartment model.}
\label{fig:2comp}
\end{figure}

\begin{example}
Consider the following two models, which have previously been shown to be indistinguishable \cite{distefano2015dynamic}:
\begin{align*}
& \textrm{Model 1} 							& & \textrm{Model 2}  \\
\dot{x}_1 &= k_{12} x_2 - (k_{21}+k_{01})x_1		& \dot{x}_3 &= k_{34} x_4 - (k_{43}+k_{03})x_3\\
\dot{x}_2 &= k_{21} x_1 - (k_{12} + k_{02})x_2 		& \dot{x}_4 &= k_{43} x_3 - k_{34} x_4 \\
y &= x_1									& y &=x_3
\end{align*}
Model 1 is shown in Figure~\ref{fig:2comp}, and Model 2 is the same model, but with $k_{02} = 0$ (rewritten with new numbering).
To show that they are input-output equivalent, we note that both models can be written in the same form by rescaling $\tilde{x}_2 = k_{12} x_2$ and $\tilde{x}_4 = k_{34} x_4$:
\begin{align*}
& \textrm{Rescaled Model 1} 								& & \textrm{Rescaled Model 2}  \\
\dot{x}_1 &= \tilde{x}_2 - (k_{21}+k_{01})x_1				& \dot{x}_3 &= \tilde{x}_4 - (k_{43}+k_{03})x_3\\
\dot{\tilde{x}}_2 &= k_{12} k_{21} x_1 - (k_{12} + k_{02})\tilde{x}_2 	& \dot{\tilde{x}}_4 &= k_{34} k_{43} x_3 - k_{34} \tilde{x}_4 \\
y &= x_1									& y &=x_3
\end{align*}
We also note that the rescaled models are each input-output equivalent to their original forms by the same logic as in Example~\ref{ex:rescale}. Using this rescaling, we see that for the models to be equivalent, we must have $k_{34} = k_{12} + k_{02}$, $k_{12} k_{21} = k_{34} k_{43}$, and $k_{21}+k_{01} = k_{43}+k_{03}$, and for initial conditions $x_3(0) = x_1(0)$ and $\tilde{x}_2(0) = k_{12} x_2(0) = \tilde{x}_4(0) = k_{34} x_4(0)$. We can rewrite this as: $k_{34} = k_{12} + k_{02}$, $k_{43} = k_{12} k_{21}/k_{34}$, and $k_{03} = k_{21}+k_{01} - k_{43}$, with initial conditions $x_3(0) = x_1(0)$ and $k_{12} x_2(0)/k_{34}$. Alternatively, we could find the same parameter relationships by calculating the input-output equations for both models and setting them equal to one another. 

Appending these equations to Model 2 above allows us to parameterize Model 2 using the same parameters as Model 1, so that they will have the same output $y$ for a given set of parameter values (i.e. values for $k_{12}$, $k_{21}$, $k_{01}$, and $k_{02}$) and initial conditions ($x_1(0)$ and $x_2(0)$), making Model 2 input-output equivalent to Model 1.
\end{example}


Using the definition of input-output equivalence, we note the following simple but useful observation:

\begin{lemma}\label{niceobs}
	Models that are input-output equivalent have the same identifiability---i.e., the same parameters are (globally or locally) identifiable or unidentifiable, and they have the same identifiable parameter combinations.
\end{lemma}

\begin{proof}
Any two parameter sets which generate the same set of outputs $\set{(\mb{y})}$ in one model must necessarily do so in the other model (because the two models are input-output equivalent). Additionally, any two parameter sets which generate the same specific output $\bf y$ in one model must necessarily do so in the other model (for some choice of initial conditions for the unobserved internal variables $\bf x$). Similarly, any two parameter sets which generate distinct output trajectories in one model will generate the same distinct trajectories in the other. Thus their identifiability properties are exactly the same (i.e. they have the same identifiable combinations and (globally or locally) identifiable or unidentifiable parameters). Put another way, because input-output equivalent models preserve the relationship between the parameters and output variables, they have the same model map and thus the same identifiabilty.
\end{proof}

We note that Lemma~\ref{niceobs} does not depend on the methods of the differential algebra approach (such as characteristic sets), or  on the model being in any specific form (i.e. it need not be in the first order, state space form given in Eq.~\eqref{eq:modelsetup}), but is rather just a direct consequence of the definition of input-output equivalence.

From Lemma~\ref{niceobs}, we can show several very simple but sometimes convenient results, described in the sections below. Some of the results below can also be shown directly fairly easily, however we include them to illustrate how they are all related to the idea of input-output equivalence. We will then examine how Lemma~\ref{niceobs} can be used to explore alternative approaches to generating input-output equations.

\subsection{Higher-order ODE models}\label{sec:higherorder}
We start with a relatively simple example, that of higher order ODE models. As noted above, the differential algebra approach generally defines models as being in state space form using first order ODEs (as in Eq.~\eqref{eq:modelsetup}), as this ensures the primacy of the resulting differential ideal \cite{Audoly2001,Ljung1994}. However it would seem natural to examine higher order ODE models, by rewriting them as a system of first order equations (for single ODEs this will be in state-space form, wherein right hand sides derivative-free, and for general systems of multivariable higher order ODEs, either in state-space form or possibly simply as first order ODEs in any form). 

Converting higher order ODEs to first order ODEs is typically done by setting each derivative to be a new variable, i.e. if we have an $n$th order differential equation of the form 
$$z^{(n)} =  h(t,z,\dot{z}, \ddot{z}, \dots, z^{(n-1)})$$ 
(potentially also a function of any inputs $\bf u$ and parameters $\bf p$), we can write it as $n$ first order equations by letting $x_1  = z$, $x_2 = \dot{z}$, \dots, $x_n = z^{(n-1)}$, and rewrite the system as 
\begin{equation*}
\begin{aligned}
\dot{x}_1 &= x_2,\\
\dot{x}_2 &= x_3,\\
 &\vdots\\
 \dot{x}_n &= h(t,x_1,x_2, \dots, x_n).
\end{aligned}
\end{equation*}
A similar transformation can be used for multivariable systems, defining new variables to account for each derivative of each variable, up to the highest derivative present in the full system of equations (although we note that in general, this could result in two derivatives appearing in the same equation, if one equation contains the highest derivative form for two variables).

Similarly, the measurement equations can be transformed to write the outputs $\bf y$ in terms of the new, first order state variables. Let us denote the higher order version of the model by $H$ and the first order version by $M$. 
$H$ and $M$ are clearly input-output equivalent (as the initial conditions directly translate from one system to the other, they have the exact same parameterization, and they give the same resulting solutions for $z = x_1$ and its derivatives). Then by Lemma~\ref{niceobs} they have the same identifiability properties (i.e. the same (locally or globally) identifiable parameters and identifiable combinations). Thus, if we wish to investigate the identifiability of a higher order ODE system, we can write it in first order form and the resulting identifiability information is the same---and if the resulting first order model is in state-space form, the usual methods of differential algebra can be applied, with primacy of the differential ideal ensured. 


\subsection{Model transformations, rescaling, and nondimensionalization}
Models can often be written in alternative forms, e.g. via nondimensionalization, and it is common to use these alternative forms to establish identifiability properties of the original model (e.g. \cite{eisenberg2013identifiability, walch2016parameter, kao2018practical, meshkat2014identifiable, vajda1989similarity}, among others).  In particular, a common tool for showing that a particular model is unidentifiable is to rescale the original model in a way that preserves the input-output relationships and then note that the rescaled model has fewer parameters than the original. We can state this idea as follows (here using rational functions of our parameters and variables, although broader classes of functions will also work):

\begin{proposition}\label{th:rescale} Let $M(t\bf u, x, y, p)$ be a model, and let $\tilde{M}$ be a tranformed version of the model, with parameters and state variables that are given by rational functions $\varphi_1(\mb{p}), \dots, \varphi_k(\mb{p})$ for the parameters, and $\phi_1(\mb{p,x}), \dots, \phi_m(\mb{p,x})$ for the variables. Suppose that the $\varphi_i$ and $\phi_i$ are such that, for a given $\bf u$ (if the model includes inputs), transforming $\bf p$ and $\bf x(0)$ yields the same output for $\tilde{M}$ as for $M$ (namely $\bf y$), and that $k < n_p$, the number of parameters of $M$. Then $M$ is not identifiable, and the identifiable combinations of $M$ can be written in terms of the $\varphi_i$. 
\end{proposition}

\begin{proof}
Suppose that we have the conditions state above, so that if we take a particular set of values for the parameters $\bf p^*$ and initial conditions $\bf x(0)^*$, transform them according to the $\varphi_i$ and $\phi_i$, and use them in $\tilde{M}$, then for a given input $\bf u$, the resulting output $\bf\tilde{y}$ of $\tilde{M}$ is equal to the output $\bf y$ for $M$. 
Let us denote the model $\tilde{M}$, together with the transformation equations $ \varphi_1(\mb{p}), \dots, \varphi_k(\mb{p})$ and $\phi_1(\mb{p,x}), \dots, \phi_m(\mb{p,x})$ by $N$ (i.e. this is $\tilde{M}$ now parameterized in the same way as $M$). $N$ is input-output equivalent to $M$. 

Since $k$ is less than $n_p$, $N$ must be unidentifiable, as the system of equations $\varphi_1(\mb{p}) = a_1, \dots$, $\varphi_k(\mb{p}) = a_k$ (for any given set of values $a_1, \dots, a_k$ for the $\varphi_i$) is underdetermined, meaning there are an infinite number of ways to generate the same values for the $\varphi_i$. Since $N$ is input-output equivalent to $M$, $M$ is also unidentifiable. 

Moreover, any identifiable coefficients of $M$ must be in terms of the $\varphi_i$. We can see this as follows: consider a set of coefficient identifiable input-output equations of $\tilde{M}$ (e.g. generated by taking its characteristic set), denoted by $\bf \tilde{\Psi}$. The coefficients of $\bf \tilde{\Psi}$ will be in terms of the $\varphi_i$, the parameters for $\tilde{M}$. Note also that $\bf \tilde{\Psi}$, when viewed as a model, is input-output equivalent to $\tilde{M}$. Let $\tilde{N}$ be the model given by the set of equations $\bf \tilde{\Psi}$, $ \varphi_1(\mb{p}), \dots, \varphi_k(\mb{p})$, and $\phi_1(\mb{p,x}), \dots, \phi_m(\mb{p,x})$. $\tilde{N}$ is input-output equivalent to $N$, and thus to $M$, but has identifiable combinations in terms of the $\varphi_i$, implying that $M$ does as well.
\end{proof}

Example~\ref{ex:rescale} illustrates this idea---by Proposition~\ref{th:rescale}, rescaling Model 1 to Model 2 implies that Model 1 is unidentifiable, as Model 2 has fewer parameters than Model 1.

\begin{figure}
\begin{center}
\includegraphics[width=0.5\textwidth]{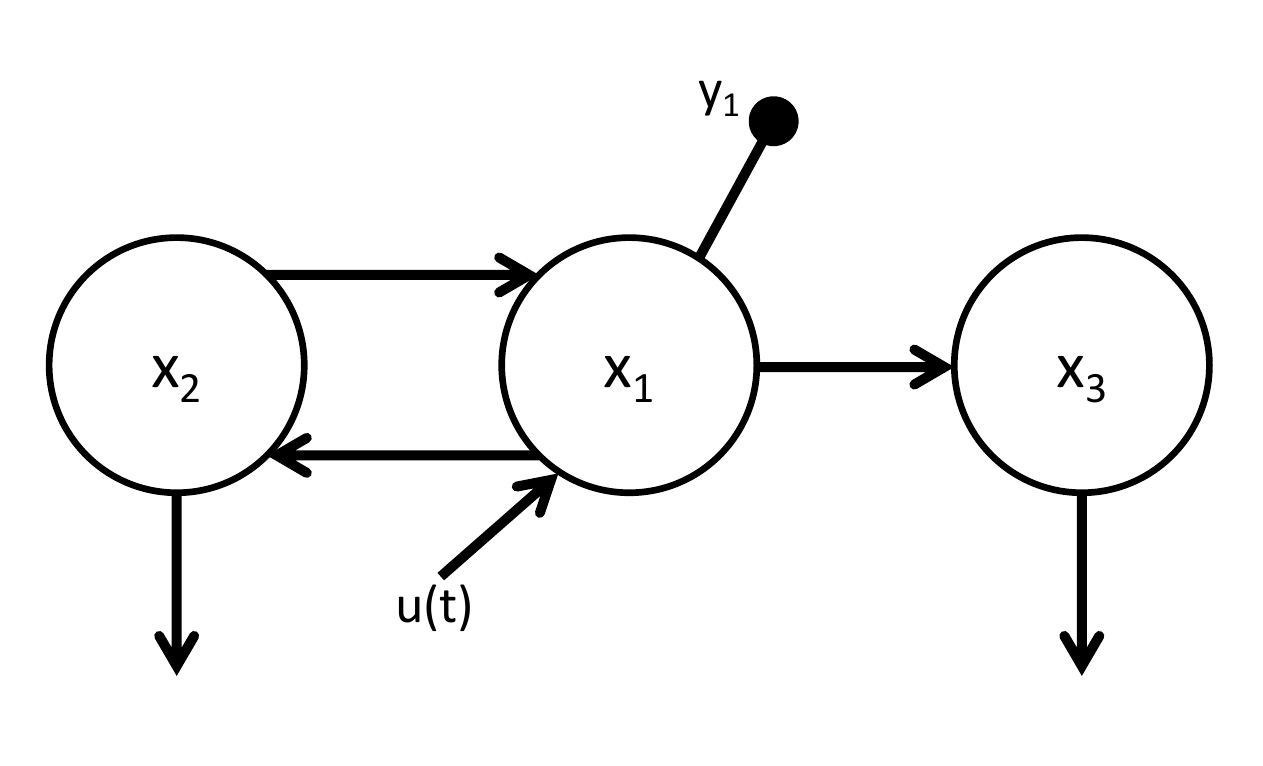}
\caption{Example of a reducible model. Because there is no connection from $x_3$ to $x_1$, $x_3$ can be removed from the model without affecting $x_1$ or the output $y$. Any parameters which appear only in the $x_3$ equation will thus be unidentifiable.}
\label{fig:reducible}
\end{center}
\end{figure}

Proposition~\ref{th:rescale} is also related to the identifiability of \emph{reducible models} \cite{distefano2015dynamic}, namely models which include variables that could be removed from the model without affecting the dynamics of the model output. An example of a reducible model is shown in Figure~\ref{fig:reducible}. Let $M$ be a reducible model. We can take the $\phi_i$ to simply drop the unnecessary variables of $M$ (and otherwise have the $\phi_i$ be the identity function), and similarly make the $\varphi_i$ drop any parameters that only appear in the equations for the reducible variables (and otherwise have the $\varphi$ be the identity function). Now $\tilde{M}$ has fewer parameters and variables, but is input-output equivalent to $M$, implying that any parameters that were removed were unidentifiable (which is clear in any case, as they do not affect the output). 

\subsection{A modular approach to identifiability}\label{sec:modular}
Additionally, Lemma~\ref{niceobs} can be used to examine when the identifiability of a model can be decomposed into submodels. Many real-world models are quite large (and so potentially computationally intractable for determining identifiablity). This raises the possibility that one might be able to decompose such large models into smaller submodels, evaluate the identifiability of each, and then combine the results to understand the identifiability of the full model. We will show here that this is in some cases possible, when the model is input-output equivalent to one wherein the connections between the submodels are measured/known.

\begin{figure}
\begin{center}
\includegraphics[width=0.42\textwidth]{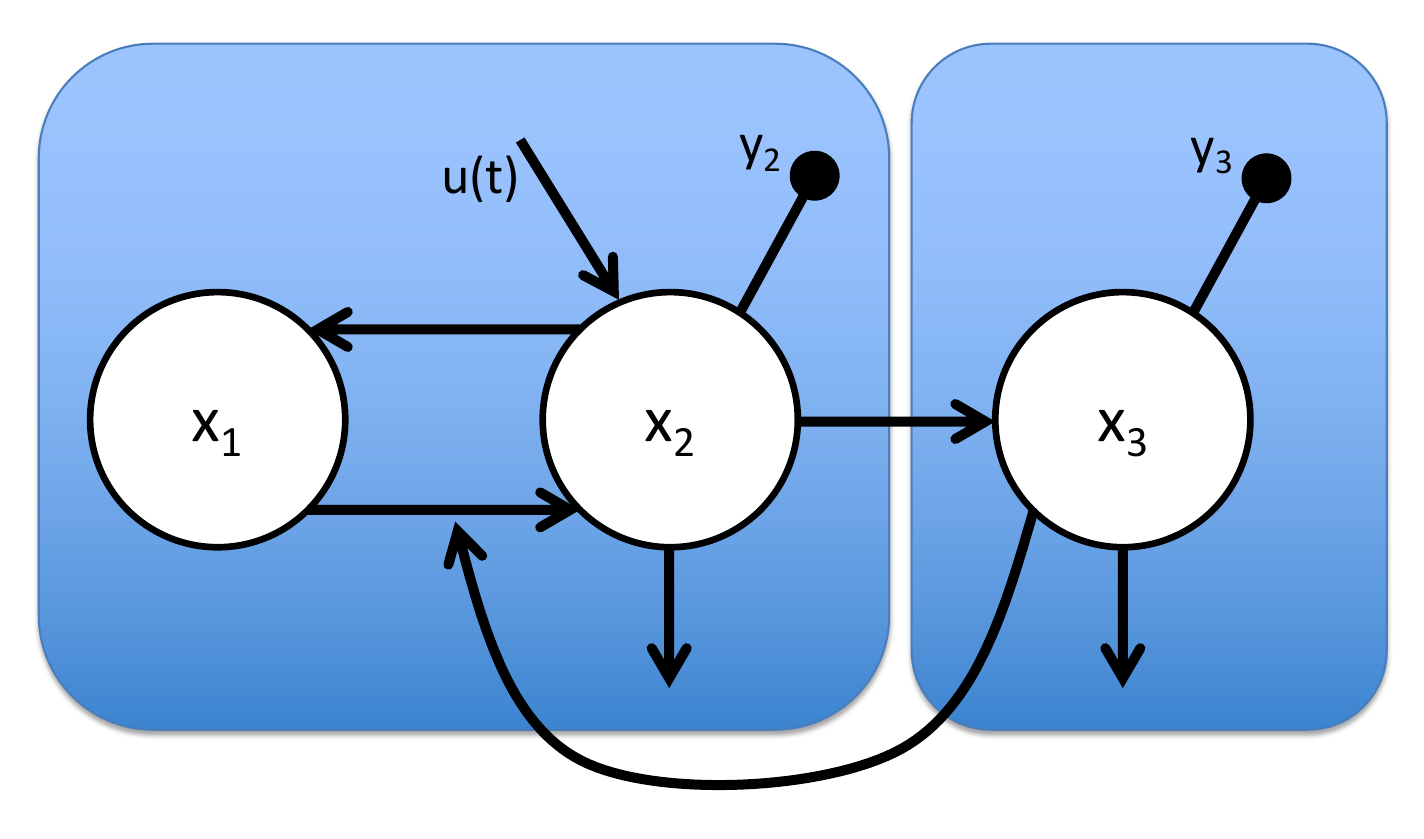}\\ \ \\
\includegraphics[width=0.3\textwidth]{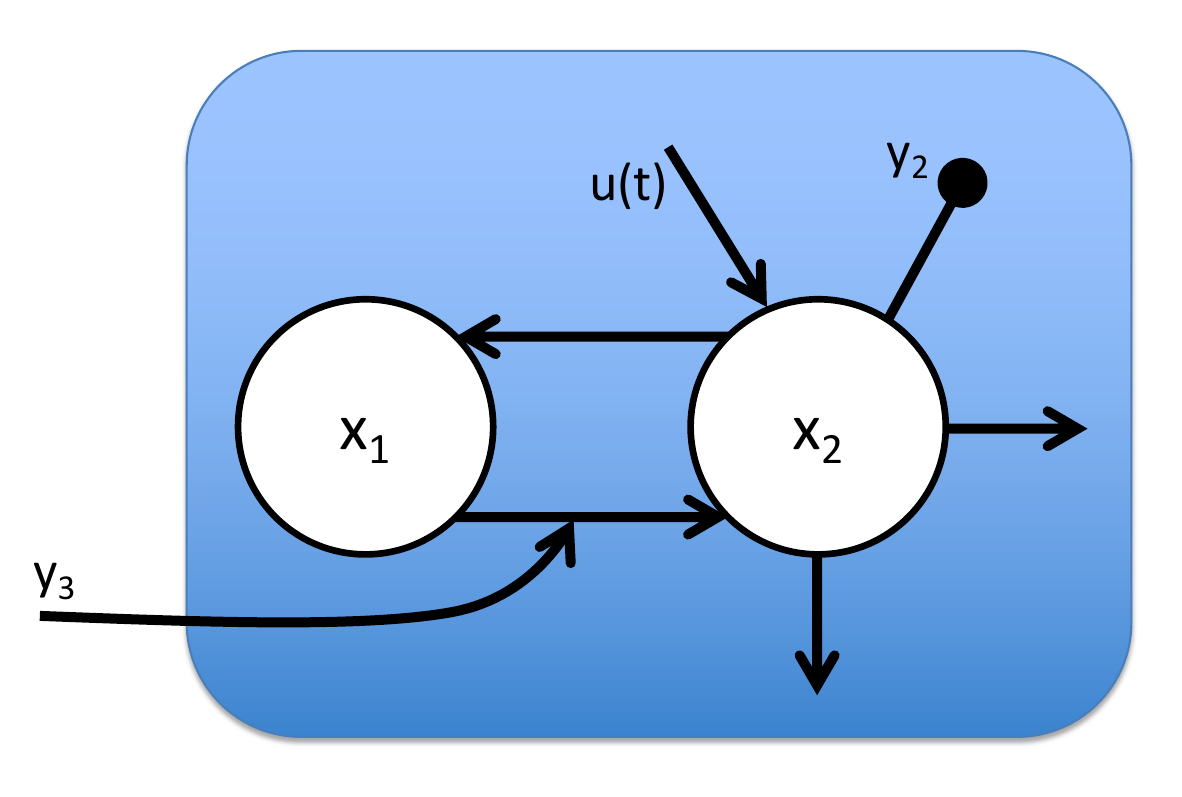} \qquad \qquad \includegraphics[width=0.17\textwidth]{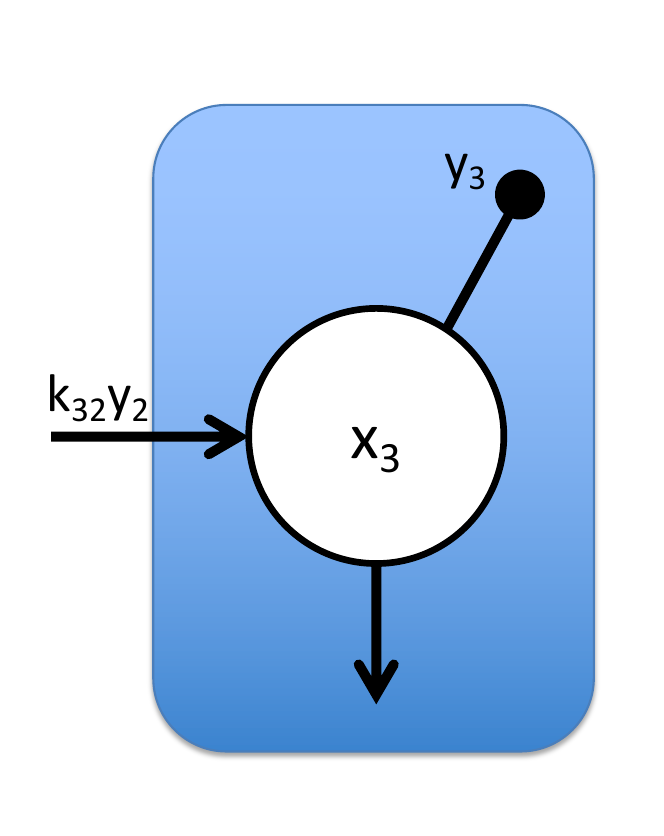}
\caption{Top row: the model given in Example~\ref{ex:modular}, illustrating how it can be broken up into two submodels, $M_2$ containing the $x_1$, $x_2$, and $y_2$ equations, and $M_3$ containing the $x_3$ and $y_3$ equations. Because $y_2 = x_2$ and $y_3 = x_3$, the connections between the two submodels can be written as $k_32 y_2$ (from $M_2$ to $M_3$ and $y_3$ (from $M_3$ to $M_2$). The second row illustrates these two submodels broken into two independent submodels, with the connections drawn as input forcing functions for each submodel.}
\label{fig:modular}
\end{center}
\end{figure}

\begin{example}\label{ex:modular} First, let us give a motivating example. Let $M$ be the following nonlinear system:
\begin{equation}
\begin{aligned}
\dot{x}_1 &= k_{12} x_2 - k_{21} x_3 x_1 \\
\dot{x}_2 &= u + k_{21} x_3 x_1 - (k_{12} + k_{32} + k_{02}) x_2\\
\dot{x}_3 &= k_{32} x_2  - k_{03} x_3 \\
y_2 &= x_2\\
y_3 &= x_3
\end{aligned}
\end{equation}
shown in Figure~\ref{fig:modular}. In this model a substance (e.g. a protein) converts reversibly between two forms ($x_1$ and $x_2$), undergoing a further conversion to state $x_3$, which then feeds back to affect the rate at which $x_1$ converts to $x_2$. We can show using the standard differential algebra approach that this model is fully identifiable. However, notice that the model can also be broken into two submodels, which are only connected by known variables. This can be accomplished by placing the $x_1$, $x_2$, and $y_2$ equations in one submodel (denoted $M_2$), and the $x_3$ and $y_3$ equations in another submodel (denoted $M_3$), as illustrated in Figure~\ref{fig:modular}. These can be written as two distinct models, each of which receives a function of either $y_2$ or $y_3$ as a known `input': 
\begin{align*}
& M_2 												& 	& M_3  \\
\dot{x}_1 &= k_{12} x_2 - k_{21} y_3 x_1 						& \dot{x}_3 &= k_{32} y_2  - k_{03} x_3 \\
\dot{x}_2 &= u + k_{21} y_3 x_1 - (k_{12} + k_{32} + k_{02}) x_2	& \\
y_2 &= x_2											& y_3 &= x_3
\end{align*}
$M_2$ and $M_3$ together (denoted $(M_2,M_3)$) are input-output equivalent to $M$ (as they are the same model, just with the $y_i$ replacing the $x_i$), but we can now simulate $M_2$ or $M_3$ independently, for a given set of inputs and outputs $u$, $y_2$, and $y_3$. We note that in each submodel, the `input' $y_i$ (i.e. $y_3$ for $M_2$ and $y_2$ for $M_3$) is not free to vary as they might be for a general input-output model---these are prespecified trajectories, which is why we do not replace them with additional $u_i$'s. Since $M_2$ and $M_3$ are now separate, we can evaluate their identifiability individually (as neither model affects the other). If we do so, we find that $M_2$ is unidentifiable, while $M_3$ is identifiable. Our identifiable combinations are:
\begin{align*}
&M_2 \textrm{\,(unidentifiable)}		& M_3 \textrm{\,(identifiable)}	  \\
&k_{21},\, k_{12},\, k_{02}+k_{32}					& k_{32},\, k_{03} 
\end{align*}
Then we see that while $M_2$ is unidentifiable by itself, once the information from $M_3$ is added (namely, that $k_{32}$ is identifiable), the combined model $(M_2,M_3)$ is identifiable, matching the results for $M$.
\end{example}

To generalize this idea, let $M(t,\bf u, x, y, p)$ be a model whose state variables $\bf x$ and outputs $\bf y$ can be partitioned into submodels $M_1, \dots, M_n$ (so that each $M_i$ is a collection of state variable equations and measurement equations, the union of which across all the $M_i$ is just $M$). 
We will call this a \emph{decomposition} of $M$ into submodels. When a variable or variables from submodel $M_i$ appear in the equations for submodel $M_j$, we will say that $M_i$ is connected to $M_j$, and call the term(s) in which these variables appear in the $M_j$ equations the connection from $M_i$ to $M_j$. Then if each parameter only appears in the equations of one submodel (so that we can partition the parameters among the submodels), we have the following:

\pagebreak
\begin{proposition}\label{th:modular1}
Let $M(t,\bf u, x, y, p)$ be a model that is decomposable into submodels $M_1, \dots, M_n$ wherein the connections between all submodels are written in terms of $t$, $\bf u$, $\bf y,$ and $\bf p$. Suppose that each parameter only appears in one submodel of $M$. Then $M$ is identifiable if and only if all submodels are identifiable. $M$ is globally identifiable if and only if all submodels are globally identifiable, and locally identifiable if and only if all submodels are identifiable but at least one submodel is only locally so. Moreover, the identifiable combinations of $M$ are precisely the union of the identifiable combinations of the $M_i$ when evaluated independently. \end{proposition}

\begin{proof} Let $M$ be a model that is decomposable into $M_1, \dots, M_n$ as described, and let us denote the set of incoming connections to submodel $M_i$ as $\chain_i$, where $\chain_i$ is a list of terms from other submodels appearing in $M_i$, with the entries of $\chain_i$ only in terms of $\bf t, u, y,$ and $\bf p$. 
We note that when estimating the parameters for $M$ from a given data set $\bf y$, we could estimate the parameters for any particular $M_i$ entirely independently, by plugging in the appropriate $y_j$'s from other submodels into $\chain_i$ as an input (i.e. as forcing function inputs in terms of $\bf t, u, y$), and then estimating the parameters from $M_i$ from the outputs whose measurement equations are contained in $M_i$. We will write $M_i$ as $M_i(t,\bf u, y, x_i, y_i, p_i)$ to indicate that $M_i$ receives some forcing function inputs in terms of $\bf t, u, y$, and has its submodel-specific variables and parameters $\bf x_i, y_i, p_i$ (noting that we can partition $\bf x, y$, and $\bf p$ into the $\bf x_i, y_i, p_i$ of each submodel since each parameter appears in only one submodel). Then let us examine how the identifiability of the $M_i$ relates to that of $M$. 

Suppose that two or more distinct points in parameter space all yield the observed data $\bf y$ when plugged in to $M$, given the specified inputs $\bf t, u, y$. 
Then for each $M_i$, these points all generate $\bf y_i^*$ as well. Additionally, since these points in parameter space are distinct, for each pair of them $\bf p^*$ and $\bf \hat{p}$, there must be some $i$ for which $\bf p^*_i \neq \bf \hat{p}_i$. Thus, if $M$ is unidentifiable, then so is at least one $M_i$, and if $M$ is locally identifiable, then those same local parameter solutions will also yield the same output in $M_i$ (and be distinct points in at least one $M_i$).

Now again suppose that we have a particular set of known data $\bf t, u, y$, and we are now estimating the parameters of a single submodel $M_i$. Suppose that two or more distinct points in the parameter space of $M_i$ yield the same observed data $\bf y_i$, given the specified inputs $\bf t, u, y$. Also take $\bf p$ to be a point in parameter space which generates $\bf t, u, y$ from $M$ (such a point exists as we are assuming the data comes from $M$). Now, for each point $\bf p^*_i$ for which $M_i$ generates $\bf y_i$, append the remaining parameters for the rest of $M$ from $\bf p$. This combined vector of parameter values will generate $\bf y$, as $\bf p^*_i$ generates $\bf y_i$ when plugged in to $M_i$ and each submodel can be simulated entirely independently once the inputs are specified as $\bf t, u, y$. Then any distinct points which generate the same output in one of the $M_i$ can be used to find distinct points which generate $\bf y$ using $M$. Thus, if any $M_i$ is unidentifiable, so will be $M$, and if any $M_i$ is locally identifiable, then those same local parameter solutions (expanded to a full parameter vector) will also yield the same output in $M$. Because any points in parameter space which yield the same output for the submodels precisely correspond with those for the full model (and vice versa from above), the identifiable combinations for the full model are the union of the combinations from the submodels.
\end{proof}

Alternatively in the more general case where we may parameters appearing in multiple submodels (as in Example~\ref{ex:modular}, where $k_{32}$ appears in both submodels), we have:

\begin{proposition}\label{th:modular2}
Let $M$ be a model that is decomposable into submodels $M_1, \dots, M_n$ wherein the connections between all submodels are written in terms of $\bf t, u, y,$ and $\bf p$. Then the identifiability of $M$ can be decomposed into that of the submodels, in the sense that we can evaluate the identifiable combinations for each submodel individually and examine the union of these identifiable combinations to determine the identifiability of the full model. Additionally, if all submodels are separately identifiable, so is the full model. 
\end{proposition}

\begin{proof} Let $M$ be a model that is decomposable into $M_1, \dots, M_n$ as described. As in the proof for Proposition~\ref{th:modular1}, if two or more distinct points in parameter space all yield the observed data $\bf y$ when plugged in to $M$, given the specified inputs $\bf t, u, y$, then those points will also do so for the $M_i$ (dropping the unnecessary parameters for each $M_i$), and will be distinct points for some $M_i$. This implies that if $M$ is unidentifiable then there is at least one $M_i$ which is also unidentifiable, or by converse that if all $M_i$ are identifiable then so is $M$.

Now, suppose that we are estimating the parameters of a single submodel $M_i$, and that two or more distinct points in the parameter space of $M_i$ yield the same observed data $\bf y_i$, given the specified inputs $\bf t, u, y$. These points in parameter space, when taken together, describe the identifiable combinations or local solutions of the parameter space for $M_i$.

As in the previous proof, take $\bf p$ to be a point in parameter space which generates $\bf t, u, y$ from $M$ (such a point exists as we are assuming the data comes from $M$). For each of our points $\bf p^*_i$ for which $M_i$ generates $\bf y_i$, append the remaining parameters for from $\bf p$. However, because the parameters of $M_i$ may also appear in other submodels, so we do not know if the $\bf p^*_i$ will give the appropriate $\bf y$ when plugged in to the other submodels. If they do, then the same argument as for Proposition~\ref{th:modular1} applies.

If not, then there exist some submodel(s) for which these points do not yield $\bf y$ (or the appropriate piece of $\bf y$) when we plug in the parameter vector $\bf p^*_i$ (in extended form). Let one such submodel be $M_j$. Then for $M_j$, the identifiable combinations and/or local solutions of the $\bf p_j$ do not match those of $M_i$ and have a distinct form. When estimating the parameters for each $M_i$ individually, if our goal is to estimate the parameters for $M$ then we would keep only those estimates which generate $\bf y$ for all submodels, i.e. the intersections of the solutions in parameter space across all submodels. This corresponds to finding common solutions for all combinations from each submodel, i.e. to evaluating the identifiability based on all submodel coefficients taken together.
\end{proof}

\begin{remark}
We note that most models are not initially written such that the connections between submodels are the output variables (as output variables are not typically written among the model equations. However, rewriting them in this way preserves input-output equivalence, so that by Lemma~\ref{niceobs}, Propositions~\ref{th:modular1} and \ref{th:modular2} are also true for models which can be rewritten in this form. 
\end{remark}


\section{Alternative approaches to generating input-output equations}
Lastly, we address one of the main goals of this paper, to examine how we can use more general methods to find input-output equations that inform identifiability analyses.

\subsection{Generalized input-output equations} As discussed in Section~\ref{sec:charset}, input-output equations are typically generated via characteristic sets, and form a set of $n_y$ monic differential polynomials only in terms of only the known (measured) variables $\bf y$ and $\bf u$, along with the unknown parameters $\bf p$, which are an implicit form of the model map $\Phi$ \cite{Audoly2001}, meaning that for a given set of parameters $\mb p$ and inputs $\mb u$, they generate the same set of outputs $\set(\mb y)$ as the original model \cite{Audoly2001}. Thus, the input-output equations are input-output equivalent to the original model, but in some sense a minimally so, because their solutions are precisely the input-output pairs of the system (with no other extraneous variables). 

However, anecdotally it is common to see that many different methods can reduce the full model system to an equivalent one in terms of only the known variables, suggesting that the (or at least a) critical feature of the input output equations is not that they are generated via characteristic sets, but rather that by eliminating the unobserved variables, the input-output equations represent an implicit form of the model map $\Phi$. 
Thus, in principle, input-output equations could be generated by a range of methods, e.g. simply by \textit{ad hoc} substitution and differentiation, using Gr\"{o}bner bases (provided we differentiate the equations sufficiently many times; as shown in \cite{Meshkat2012}), or via various types of differential Gr\"{o}bner bases (such as \cite{Mansfield1991}).

To account for these other potential approaches, we will generalize our definition of the input-output equations as follows:
\begin{definition}
A set of \emph{generalized input-output equations} for a model $M$ is defined as any set of $n_y$ monic differential polynomials that (when set equal to zero) are input-output equivalent to $M$ and are in terms of only the known variables $\bf y$ and $\bf u$, their derivatives, and the parameters $\bf p$. 
\end{definition}

We note that because the input-output equations generated by the characteristic set embody the model map $\Phi$, they fit this definition. Initial conditions are often treated less explicitly when considering the input-output equations, but we note that the rest of the characteristic set provides a translation map from the initial conditions of $\bf y$ to those of $\bf x$ \cite{Bellu2007, Audoly2001}.
We also remark that the definition above can be generalized somewhat further,  e.g. to allow for integrals of $\bf y$ and $\bf u$ (with appropriate consideration of initial conditions), a topic to be explored in future work.

\begin{remark} However, a note of caution regarding input-output equivalence when using substitution and other methods to calculate generalized input-output equations: one must be careful not to cancel any terms containing variables ($\bf x, y$, or $\bf u$) from both sides of the equations (e.g. when using substitution). Doing so will change what solutions the resulting input-output equations represent, as by cancelling these terms one is therefore assuming the cancelled terms are not zero (thus eliminating some potential solutions to the model). This effectively changes the manifold of solutions to the system and makes the resulting equations no longer input-output equivalent to the full model. 

This cancelling property can be useful when one is deliberately exploring specific families of solutions to the model (e.g. solutions to the model for specific initial conditions as in \cite{Saccomani2003}), but does not maintain input-output equivalence for the full system. 
\end{remark}

\subsection{Coefficient identifiability} 
In addition to being input-output equivalent to the original model, the other key property of input-output equations generated via characteristic sets is that the coefficients of the characteristic set input-output equations are identifiable and contain all identifiability information for the model, meaning that to test the injectivity of the model map, we need only test the injectivity of the map from $\bf p$ to the coefficients $\bf c(p)$. In the general case, we will call such input-output equations \emph{coefficient-identifiable} input-output equations. 

One might expect that generalized input-output equations will always be coefficient-identifiable, and indeed we will show this to be true in the single-output case. However, for multiple output systems, we have more than one input-output equation, raising
the possibility of playing the equations against one another to generate spurious coefficients that give false identifiability results. For example, suppose that we have a set of coefficient-identifiable input output equations generated from the characteristic set:
\begin{equation}\label{eq:ipop}
	\begin{aligned}
		\Psi_1(\mathbf{y}, \mathbf{u}, \mathbf{p}) &=0\\
		\Psi_2(\mathbf{y}, \mathbf{u}, \mathbf{p}) &=0
	\end{aligned}
\end{equation}
and suppose that these input-output equations show that the parameter $p_1$ is unidentifiable. 
Then we can generate an alternative form of input-output equations:
\begin{equation}\label{eq:badipop}
	\begin{aligned}
		\Psi_1(\mathbf{y}, \mathbf{u}, \mathbf{p}) + p_1\Psi_2 &=0\\
		\Psi_2(\mathbf{y}, \mathbf{u}, \mathbf{p}) &=0
	\end{aligned}
\end{equation}
Because $\Psi_2$ is monic, and $\Psi_1$ does not contain its leader (since $\Psi_1$ and $\Psi_2$ are reduced with respect to each other), the first equation of Eq.~\eqref{eq:badipop} will have a monomial with a coefficient $p_1$, making it appear that $p_1$ is identifiable when we know it is not. Eq.~\eqref{eq:badipop} is a generalized input-output equation, because it is input-output equivalent to Eq.~\eqref{eq:ipop} (and thus to the model), but it is not coefficient-identifiable.

\inlinesec{General coefficient-identifiable input-output equations} To address this, we will assume that when we have multiple outputs, our set of $n_y$ generalized input-output equations are fully autoreduced (based on some ranking among the $\bf y$ and $\bf u$), in the sense that using the Ritt pseudodivision algorithm (as in \cite{Audoly2001}) will not change the set of equations (although we do not require the equations to be generated via Ritt's pseudodivision or any other characteristic set method). This does not necessarily mean that our generalized input-output equations are themselves generated via a characteristic set of the model (as we will see in some of the examples), but rather that they are in some sense a characteristic set of themself (discussed further below). We note that if $n_y = 1$, then because there is only one input-output equation, it is already trivially autoreduced. This condition is stronger than is strictly needed (as discussed below), but we will show it is sufficient to ensure coefficient-identifiability.

In practice, once one has a set of generalized input-output equations, it is often easy to either verify that they are reduced with respect to one another (as is typically the case in practice when using substitution to calculate the input output equations) or perform the reduction (which may be simpler as we then only have $n_y$ equations rather then the full set of equations for the variables, constraints, and outputs). 



With these assumptions, Lemma~\ref{niceobs} lends itself nicely to our goal of finding other or more general methods of calculating coefficient-identifiable input-output equations. 
However, first we must address some details of working with input-output equations when they are written in state-space form. 
While we already know from Section~\ref{sec:higherorder} above that the identifiability of the original model, the input output equations, and the input-output equations in first order form all have the same identifiability properties (i.e. same globally and/or locally identifiable parameters, same identifiable combinations), we will need to understand 1) whether the first order form of the input output equations is in state space form and 2) how the characteristic set behaves when the input-output equations are transformed this way---in particular, we will show that taking a characteristic set simply unravels the transformed, first order equations to return the input output equations in their original form.

\begin{lemma}\label{th:ipopcharset} Let $\bf \Psi$ be a set of generalized input-output equations that are fully autoreduced with respect to some ranking. Then $\bf \Psi = 0$ can be written in state-space form (as in Eq.~\eqref{eq:modelsetup}). Additionally, let $M$ be the model given by $\bf\Psi = 0$ written in state-space form. Then the input-output equations generated using a characteristic set of $M$ are precisely $\bf \Psi$.
\end{lemma}

The proof of this Lemma~\ref{th:ipopcharset} follows by writing the input-output equations in first-order form, and then using the fact that they are autoreduced to show that they are in state-space form (as no other equation can contain the leader of each equation). We then note that the usual ranking on the variables \cite{Audoly2001} ensures that the standard reduction algorithm using Ritt's pseudodivision will return the original input-output equations. The proof is fairly straightforward, but somewhat long, and so is included in Appendix~\ref{sec:lemmaproof}. 


\begin{theorem}\label{th:genID}
	Any generalized input-output equation for a model with a single output ($n_y = 1$) is a coefficient-identifiable input-output equation. If we have multiple outputs, then if a set of generalized input-output equations for the model is fully autoreduced, it is a set of coefficient-identifiable input-output equations. 
\end{theorem}

\begin{proof}
%
Suppose we have a set of generalized input output equations for a model $M(\mb{u, y, p})$, which we will denote as $\mb{\Psi}$ (in the single output case, $\mb{\Psi}$ is only one equation). By Lemma~\ref{th:ipopcharset}, the equations $\mb{\Psi}=0$ can be written in state-space form, forming a model we will call $N$. As noted in Section~\ref{sec:higherorder}, $N \sim \bf \Psi$, and thus $N\sim M$, making the identifiability of $N$ the same as that of the original model. To evaluate the identifiability of $N$, we take its characteristic set, denoted $char(N)$, and note that by Lemma~\ref{th:ipopcharset}, the resulting input-output equations from $char(N)$ are $\bf\Psi$. This means that $\bf\Psi$ is a characteristic set-generated input-output equation for the state-space model $N$ (i.e. for itself, viewed as a model). Thus, because characteristic set-generated input output equations are coefficient identifiable, and $N$ is input-output equivalent to the original model $M$, $\bf \Psi$ is a coefficient-identifiable input-output equation for the model.
\end{proof}

We can restate the single-output result another way: if the model has only one output, then we can test identifiabilty of the model using the coefficients of the input output equation (i.e. testing whether the map $\bf p\mapsto c(p)$ is injective), regardless of how the input-output equations were generated (e.g. via substitution, Gr\"{o}bner bases, etc.). In the multi-output case, a similar statement applies, however the set of generalized input-output equations are no longer necessarily reduced with respect to one another, meaning that the set of generalized input-output equations may not form a characteristic set of itself. 

Assuming the generalized input-output equations are autoreduced with respect to some ranking is sufficient to ensure this, but it
is a stronger condition than is necessary---in many cases, input-output equations which are not autoreduced will still be coefficient identifiable. For example, consider the following system: 
\begin{equation}
	\begin{aligned}
		\dot{x} &= -x,\\
		y_1 &= k_1 x,\\
		y_2 &= k_2 x.
	\end{aligned}
\end{equation}
There are two ways to eliminate $x$ using simple substitution, one of which is to note that $x = y_1/k_1$, resulting in the input-output equations: $\Psi_1 =  \dot{y}_1 + y_1, \Psi_2 = y_2 - \frac{k_2}{k_1} y_1$,
which are the input-output equations from the characteristic set under the ranking $y_1<y_2 <x$,
making the coefficients identifiable. 

An alternative set of generalized input-output equations for this model is: $\Psi_3 = \dot{y}_1 + y_1, \Psi_4 = y_2 - \frac{k_2}{k_1} y_1 + 5(\dot{y}_1 + y_1)$. These
are clearly not autoreduced under either ranking but are coefficient-identifiable (whereas if we replaced the $5$ with a $k_1$ they would no longer be). Most solution methods that we have observed in practice do not add any extraneous terms and thus tend to result in coefficient identifiable equations, however further work is needed to more tightly bound the conditions needed to guarantee coefficient identifiability. For instance, the proofs above will also work if we instead assume that autoreduction doesn't change the coefficients of the generalized input-output equations (but we do allow it to change the form of the equations themselves), rather than assuming the input-output equations are autoreduced.


\subsection{Approaches to generating input-output equations}

Theorem~\ref{th:genID} lets us show that several common approaches to generating generalized input-output equations will generate coefficient-identifiable input-output equations. For instance, simple substitution does not alter the model solutions, so long as care is taken not to divide by any variables or coefficients which cannot be assumed to be non-zero. Thus, this approach will preserve input-output equivalence, meaning that substitution generates generalized input-output equations. For single-output models, such input-output equations are then coefficient-identifiable. For the multi-output case, we must check that the equations are autoreduced with respect to each other, but it is often easy to simply choose a ranking/relabeling of the variables which ensures this once the input-output equations are calculated (since typically the different outputs $\bf y$ measure different model variables, resulting in different levels of differentiation or even entirely different variables in the resulting input-output equations).
 
Similarly, if we revisit the modular approach to identifiability discussed in Section~\ref{sec:modular}, we see that if we have a model $M$ which is decomposable as $M_1, \dots, M_n$, where the connections between the submodels are only in terms of $t, \bf, u, y, p$, then union of the input-output equations we generate from each submodel separately form a set of generalized input-output equations for the model. Moreover, if each submodel only includes one measurement equation, then the union of any set of coefficient-identifiable input-output equations for each submodel forms a set of coefficient-identifiable input-output equations for the full model. 

\inlinesec{Combining single-output input-output equations} To illustrate another alternative method for calculating input-output equations, let us consider the case where we have multiple outputs. One might wonder, can we calculate the input-output equations separately for each output, as though it was the only measurement, and then combine the results to give a set of input-output equations when all measurements are considered? 

To show this is true, suppose we have a model $M$ of the form in Eq.~\eqref{eq:modelsetup}, and $n_y =n>1$. Let $M_i$ be the model obtained by taking $M$ and replacing the measurement equations with the single measurement equation $y_i = g_i(\mb{x}, t, \mb{p})$, i.e. as though $y_i$ is the only measurement (note this is different from the $M_i$ submodel definition in Section~\ref{sec:modular}---here the $M_i$ includes the full set of $\bf x$ equations, and one output equation for $y_i$). Suppose we have $\Psi_1, \dots, \Psi_n$, each a coeffcient-identifiable input-output equation obtained for $M_i$ (via whatever method we choose). Then for a given input $\bf u$ and parameters $\bf p$, each $\Psi_i=0$ will yield the same output trajectories $\set(y_i)$ as $M$. Thus, the set $\Psi_1, \dots, \Psi_n$ will generate the same set of output trajectories for all of the $y_1, \dots, y_n$ as $M$ (noting that the solutions and initial conditions for each $\Psi_i$ are independent as they are each in terms of distinct $y_i$). Then the set $\Psi_1, \dots, \Psi_n$ is input-output equivalent to the original model $M$, making $\Psi_1, \dots, \Psi_n$ a set of generalized input-output equations. Moreover, because each $\Psi_i$ is in terms of only $y_i$, they cannot contain the leaders of one another, making this set fully autoreduced. Then the set $\Psi_1, \dots, \Psi_n$ forms a set of coefficient-identifiable input output equations. 

We note that this set will be distinct from the input-output equations that would be generated by taking the characteristic set of $M$, as the characteristic set input-output equations will typically contain equations with multiple $y_i$'s. Thus this approach is an example of a distinct way of calculating coefficient-identifiable input-output equations. Calculating the input-output equations by neglecting other outputs will typically require more tedious calculations than those with the full output set, but this result can be useful when one wants to compare identifiability for a range of possible combinations of inputs---we can calculate input-output equations for each, and then combine them as desired. 
We also note that we can generalize this result to the case where we partition the set of outputs and calculate the input-output equations for each subset in the partition, then combine the results.


\inlinesec{Gr\"{o}bner Bases and Differential Gr\"{o}bner Bases} As another example of methods to calculate input-output equations, let us consider Gr\"{o}bner bases and the differential Gr\"{o}bner bases of Mansfield \cite{Cox1996, Mansfield1991}. We note that Gr\"{o}bner bases have already been shown to generate coefficient-identifiable input-output equations in \cite{Meshkat2012}, but we consider them here as an example.

\begin{proposition}  Both Mansfield differential Gr\"{o}bner bases and Gr\"{o}bner bases (with suffiently many derivatives of the model equations added) of a model $M$ in the form of Eq.~\eqref{eq:modelsetup}
contain a complete set of generalized input output equations.
\end{proposition}

\begin{proof} Let us first show this for differential Gr\"{o}bner bases. Let $M$ be as in Eq.~\eqref{eq:modelsetup}, and let $\tilde{f} = \mb{\dot{x}}(t,\mb{u,p}) - f(\mb{x},t,\mb{u,p})$, and $\tilde{g} = \mb{y(x,u,p)} - g(\mb{x},t,\mb{p})$, so that the model is given by $\tilde{f} = 0$, $\tilde{g} = 0$. Let $mgb(\tilde{f},\tilde{g})$ be a Mansfield differential Gr\"{o}bner basis of the model, 
taking the usual lexicographic ordering with $\mb{u<\dot{u}<\cdots<y<\dot{y}}<\cdots$ $<\mb{x <\dot{x}}$ $<\cdots$ (as given in \cite{Audoly2001}). Then $\{mgb\} = \{\tilde{f},\tilde{g}\}$, so $V(\{mgb\}) = V(\{\tilde{f},\tilde{g}\})$ and we have that solutions to $mgb$ are precisely the allowed trajectories of the model. 
We also know that the characteristic set $char(\tilde{f},\tilde{g}) \subset \{\tilde{f},\tilde{g}\} = \{mgb\}$.  Since the characteristic set contains a set of input output equations, we have that $\{\tilde{f},\tilde{g}\} \cap \mb{R\{y,u}\} \neq \varnothing$.  Then by the differential elimination theorem of Mansfield \cite{Mansfield1991}, we have that $mgb\cap \mb{R\{y,u}\}$ is a Mansfield Gr\"{o}bner basis for the ideal $\{\tilde{f},\tilde{g}\} \cap \mb{R\{y,u}\}$, i.e. 
$$\{mgb\cap \mb{ R\{y,u\}}\} = \{\tilde{f},\tilde{g}\} \cap \mb{R\{y,u\}}.$$
Moreover, because solutions of $\{\tilde{f},\tilde{g}\} \cap \bf R\{y,u\}$ are precisely the allowed input-output trajectories, and $mgb\cap\bf R\{y,u\}$ generates this ideal, then $V(mgb\cap\bf R\{y,u\})$ is precisely all input-output trajectories. Since $mgb\cap\bf R\{y,u\}$ is parameterized in the same way as $M$, it is thus input-output equivalent to $M$, and thus $mgb$ contains a set of generalized input output equations.  A very similar argument shows that the same is true for algebraic Gr\"{o}bner bases, provided we begin by taking sufficiently many derivatives of the model equations to include all the differential variables needed to generate the characteristic set (and use the usual Gr\"{o}bner basis elimination theorem \cite{Cox1996} rather than the differential version). 
\end{proof}


\inlinesec{Examples} Next, let us illustrate some of the above methods with two examples.

\begin{example}{\textit{Single-output 2-compartment model}.} To begin, let us return to the linear 2-compartment model from Figure~\ref{fig:2comp}, which is frequently used in pharmacokinetics, and has been shown previously by several methods to be unidentifiable \cite{Meshkat2009, Audoly2001}:
\begin{equation}
\begin{aligned}
\dot{x}_1 &= u(t) + k_{12} x_2 - (k_{01} + k_{21}) x_1 \\
\dot{x}_2 &= k_{21} x_1 - (k_{02} + k_{12}) x_2 \\
y &= x_1/V
\end{aligned}
\label{eq:2comp}
\end{equation}
where $x_1$ represents the mass of a substance in the blood (e.g. a hormone or drug), and $x_2$ represents the mass of the substance in the tissue.  The drug exchanges between blood and tissues, and is degraded/lost in both compartments, at the rates given by the $k_{ij}$'s above. The function $u(t)$ represents a known input of the drug into the blood. The model output $y = x_1/V$ is the blood concentration of the drug, where $V$ is the blood volume. The $k_{ij}$'s and $V$ are unknown parameters to be estimated. 

To examine the identifiabilty of \eqref{eq:2comp}, we start by generating a set of input-output equations. Here we will instead use \emph{ad hoc} substitution to generate the input-output equations.  We start by replacing $x_1$ with $x_1 = y V$, to give:
\begin{equation*}
\begin{aligned}
\dot{y}V &= u(t) + k_{12} x_2 - (k_{01} + k_{21}) yV \\
\dot{x}_2 &= k_{21} yV - (k_{02} + k_{12}) x_2 .
\end{aligned}
\end{equation*}
Next, we solve the first equation for $x_2$, to yield $x_2 = \frac{-u(t)+k_{01} V y+ k_{21} V y+V \dot{y}}{k_{12}}$, which we plug into the second equation (differentiating to give $\dot{x}_2$) to yield:
$$k_{12}k_{21} V y + (k_{02}+k_{12}) \left(u(t)- V(k_{01}+k_{21}) y+V\dot{y}\right) + \dot{u}(t)-V(k_{01}+k_{21}) \dot{y} + V \ddot{y}$$
Collecting terms and making the polynomial monic (by dividing by the coefficient of the leading term $\ddot y$) yields the input output equation:
\begin{equation}
-\frac{k_{02}+k_{12}}{V} u(t) -\frac{\dot{u}(t)}{V}+(k_{01} k_{02}+k_{01} k_{12}+k_{02} k_{21}) y + (k_{01}+ k_{02}+k_{12}+k_{21}) \dot{y} + \ddot{y}. \label{eq:2compIPOP}
\end{equation}
We note that this is the same input-output equation that would be achieved via a characteristic set, but requires fewer steps than the characteristic set algorithm (in part because the characteristic set maintains a set of three equations, each with different leaders, which must all be reduced with respect to one another, rather than focusing on just the input-output equation).  However, this example is simple enough that both approaches take similar amounts of computational time in Mathematica (the substitution approach took slightly less time than the characteristic set, 
but the improvement was basically insignificant).

In calculating the input-output equation, the differential Gr\"{o}bner basis approach took 0.022579 seconds of CPU time in Mathematica, the substitution approach took 0.028624 seconds, and the characteristic set approach took 0.029163 seconds.  Using a standard Gr\"{o}bner basis to calculate the input-output equations took significantly longer, at 0.339387 seconds of CPU time.  

By Theorem \ref{th:genID}, the coefficients to (\ref{eq:2compIPOP}) are identifiable.  Then to test the identifiability of the individual parameters $(k_{01},k_{02},k_{12},k_{21}, V)$, we must test injectivity of the map $\bf c(p)$.  Thus, suppose we have an alternative set of parameters $(a_1, a_2, a_3, a_4, a_5)$ which also yield the same output.  As the coefficients for  the input output equation are identifiable, we have that:
\begin{equation*}
\begin{aligned}
-\frac{k_{02}+k_{12}}{V} &= -\frac{a_2+a_3}{a_5}\\
-\frac{1}{V} &= -\frac{1}{a_5}\\
k_{01} k_{02}+k_{01} k_{12}+k_{02} k_{21} & = a_1 a_2+a_1 a_3+a_2 a_4\\
k_{01}+ k_{02}+k_{12}+k_{21} &= a_1+ a_2+a_3+a_4
\end{aligned}
\end{equation*}
Solving for $(k_{01},k_{02},k_{12},k_{21}, V)$ following the algorithm in \cite{Audoly2001} reveals that the model is unidentifiable, with identifiable combinations $k_{01}+k_{21}$, $k_{02}+k_{12}$, $k_{12}k_{21}$ and one identifiable parameter, $V$ (as also shown in \cite{Audoly2001}). 
\end{example}

\begin{example}{\textit{Multi-output 2-compartment model}.} To illustrate how one can combine input-output equations calculated for individual outputs, let us consider the same model as the previous example, Eq.~\eqref{eq:2comp}, but let us now also measure the second compartment, i.e. our outputs are now $y_1 = x_1/V$ and $y_2 = x_2/V$. In this case, calculating the input output equations simply results in the model equations, now with the $x_i$'s replaced by $y_i$'s:
\begin{equation}
\begin{aligned}
\dot{y}_1 - u(t) - k_{12} y_2 + (k_{01} + k_{21}) y_1 \\
\dot{y}_2 - k_{21} y_1 + (k_{02} + k_{12}) y_2 
\end{aligned}
\label{eq:2IPOP}
\end{equation}
From \eqref{eq:2IPOP}, we can see that all four model parameters are globally identifiable. Now let us examine what happens if we calculate the input-output equations for each measurement equation on its own and then combine them.

The input-output equation for $y_1$ alone is given in Eq.~\eqref{eq:2compIPOP}. The input-output equation for $y_2$ alone is:
$$\ddot{y}_2 + \dot{y}_2 (k_{01} + k_{02} + k_{12} + k_{21}) + y_2 (k_{01} k_{02} + k_{01} k_{12} + k_{02} k_{21}) - \frac{k_{21}}{V}u$$
The coefficients of the two input-output equations are mostly the same, however, the distinct coefficients are: $-\frac{k_{02}+k_{12}}{V}, -\frac{1}{V}, \frac{k_{21}}{V}$, and the coefficients common to both equations, $k_{01} k_{02}+k_{01} k_{12}+k_{02} k_{21}$ and $k_{01}+ k_{02}+k_{12}+k_{21}$. From these coefficients, we see that we can solve for all parameters uniquely, indicating global identifiability of the model and matching the results from Eq.~\eqref{eq:2IPOP}.
\end{example}


\beginsupplement
\section{Algebra and differential algebra background}\label{sec:diffalg}

In this section, we present a very brief overview of the differential and computational algebra concepts discussed in this paper.  For full details on the fundamentals of differential and computational algebra methods, the reader is referred to \cite{Ritt1950, Cox1996, Mansfield1991}.  Let $\bf R$ be a ring in the usual algebraic sense, and $\bf x$ a set of indeterminates.   For our applications, $\bf R$ will represent the field of coefficients for an ODE model, so that we will typically consider 
$\bf R = \R(p)$ or $\C(\bf p )$, where $\bf p$ are the model parameters (alternatively some authors take $\bf R$ to be $\R$ or $\C$, with the parameters $\bf p$ simply acting as placeholders for the particular values in $\R$).  

A \emph{differential ring} is simply a ring in the usual algebraic sense, together with a differentiation operation which obeys the usual linear and product rule properties for derivatives.  For ODE models, we typically extend $\bf R[x]$ to form a differential ring with derivatives in time $t$, denoted $\bf R\{x\}$, by adding an additional derivative operation, in this case the usual polynomial derivative where we take derivatives from the ring of constants $\bf R$ to be zero.  Elements of $\bf R\{x\}$ can be thought of as elements of $\bf R[x, x', x'', \dots]$, where $\bf x'$ represents the set of derivatives of elements of $\bf x$ with respect to $t$ (where $\bf x$ is our set of variables).  We note that for convenience we will often view a particular differential polynomial in $\bf x$ as an element of $\mb{R}[\mb{x, x', \dots, x}^{(n)}]$, where is $n$ is the highest derivative of $\bf x$ appearing in the polynomial.  

Typically when working with differential polynomial rings, a ranking on the variables is chosen, in our case a ranking of the form $\mb{u<\dot{u}<\ddot{u}<\cdots<y<\dot{y}<\ddot{y}}<\cdots$ $<\mb{x <\dot{x}<\ddot{x}}$ $<\cdots$, with the rankings within each variable and derivative usually following numerical order, e.g. so that among the $\bf x$ we have $x_1 < x_2 < \cdots <x_n$ (as given in \cite{Audoly2001}).  This allows one to determine leading terms, make polynomials monic, etc.  The \emph{leader} of a differential polynomial is defined as the highest ranking derivative of that polynomial (which can be a derivative of order $0$).  Choosing a ranking allows us allows us to fix the coefficients of the input-output equations uniquely, by dividing by the coefficient of the leading term to make the polynomials monic \cite{Meshkat2009, Bellu2007}. 

Let $S$ be a set of differential polynomials in $\bf R\{x\}$.  The set of all polynomials that can be formed from elements of $S$ by addition, multiplication by elements of $\bf R\{x\}$, and differentiation is called a \emph{differential ideal} generated by $S$, which we write as $\{S\}$.  For a given set of polynomials $S$ (or differential polynomials, where we simply view the derivatives of variables as additional indeterminates), the variety $V(S)$ is defined in the usual way as the set of points for which all polynomials in $S$ are zero.  A differential ideal $I$ is called \emph{prime} if $ab\in I$ implies that either $a \in I$ or $b \in I$ and is called \emph{perfect} if $a^k \in I$ implies $a \in I$ (i.e. a perfect ideal coincides with its radical).

There are several methods for manipulating systems of polynomials and differential polynomials, including the Gr\"{o}bner basis and characteristic set methods discussed here, as well as methods of resolvents, among others \cite{Mansfield1991, Aistleitner2010}.  The usual method used to generate the input-output equations in identifiability for ODE models is the \emph{characteristic set} \cite{Ritt1950}.  
A characteristic set of a set of polynomials is defined to be a \emph{chain} of minimal rank in the differential ring, where chains of polynomials are formed by using pseudoreduction \cite{Ritt1950} to reduce the rank of the polynomials compared to one another, until a minimal, autoreduced set is reached.  For details on characteristic sets and their uses in identifiability, see \cite{Ollivier1990, Audoly2001, Bellu2007, Ljung1994}, and the overview given in \cite{Saccomani2003}. 
The pseudoreduction algorithm to generate a characteristic has been outlined in detail in \cite{Ritt1950, Audoly2001, Bellu2007}. 

Gr\"{o}bner bases are one of the most common tools in computational algebra, and hence are a natural generalization of the characteristic set approach for generating input-output equations, as numerous fast methods for calculating Gr\"{o}bner bases have been developed (e.g. Faugere algorithm \cite{Faugere1999}, Gr\"{o}bner walk methods \cite{Collart1993}).  A \emph{Gr\"{o}bner basis} of an ideal $I$ is a generating set for that ideal 
such that the remainder of any element of ring yields zero if and only if that element is an element of $I$.  For more information on Gr\"{o}bner bases, the reader is referred to \cite{Cox1996}.  When applying Gr\"{o}bner bases to differential polynomials in $\bf x$, we typically work in $\mb{R}[\mb{x, x', \dots, x}^{(n)}]$ (where $n$ is the highest order derivative appearing in our set of polynomials), where we treat derivatives of $\bf x$ as new indeterminates.  

It would be natural to extend Gr\"{o}bner basis theory to the differential case, and indeed there have been multiple formulations of differential Gr\"{o}bner bases \cite{Mansfield1991, Aistleitner2010, CarraFerro1987}.  Part of the difficulty in extending the Gr\"{o}bner bases to the differential case is in incorporating the differential structure of the ring, which lends itself to psuedoreduction rather than conventional reduction as is done in algebraic Gr\"{o}bner bases \cite{Mansfield1991, Aistleitner2010}.  Mansfield Gr\"{o}bner bases \cite{Mansfield1991} surmount this difficulty by developing a pseudoreduction formulation of differential Gr\"{o}bner bases, and so we use this formulation here.  A \emph{Mansfield differential Gr\"{o}bner basis} of a differential ideal $I$ is a generating set of $I$ such that full pseudoreduction of any element of $I$ yields zero.  Details and comparisons of Mansfield and algebraic Gr\"{o}bner bases can be found in Mansfield's thesis \cite{Mansfield1991}.

\section{Proof of Lemma~\ref{th:ipopcharset}}\label{sec:lemmaproof}

\noindent\textbf{Lemma~\ref{th:ipopcharset}} \textit{Let $\bf \Psi$ be a set of generalized input-output equations that are fully autoreduced with respect to some ranking. Then $\bf \Psi$ can be written in state-space form (as in Eq.~\eqref{eq:modelsetup}). Additionally, let $M$ be the model given by $\bf\Psi = 0$ written in state-space form. Then the input-output equations generated using a characteristic set of $M$ are precisely $\bf \Psi$.}

\begin{proof} 
Recall that we take the ranking on our variables $u_1 < u_2 <\cdots <y_1 < y_2 <\cdots < \dot{u}_1<\dot{u}_2 <\cdots< \dot{y}_1<\dot{y}_2 < \cdots < x_1<x_2 < \cdots< \dot{x}_1<\dot{x}_2 <\cdots$, and so on. 

Let us begin with the single output case, $n_y = 1$. Let our input-output equation $\Psi$ be given by $y^{(n)} = h(t,\mb{u},y,\mb{p})$, where $y^{(n)}$ is the leader of $\Psi$ and $h$ is a rational function obtained by solving for $y^{(n)}$ and moving it to the left side of the equation. Then we can write $\Psi$ as $n$ first order equations by letting $x_1  = y$, $x_2 = \dot{y}$, \dots, $x_n = y^{(n-1)}$, and rewrite the system as 
\begin{equation*}
\begin{aligned}
\dot{x}_1 &= x_2,\\
\dot{x}_2 &= x_3,\\
 &\vdots\\
 \dot{x}_{n-1} &= x_n,\\
 \dot{x}_n &= h(t,\mb{u}, x_1,x_2, \dots, x_n, \mb{p}),\\
 y &= x_1,
\end{aligned}
\end{equation*}
which we note is in state space form as in Eq.~\eqref{eq:modelsetup}. We also note that we did not transform the derivatives of the inputs (if any), but this can be addressed by defining derivatives of the input variables as new inputs. Following the usual differential algebra method, we rewrite this system as a set of differential polynomials as follows (clearing denominators of the $\dot{x}_n$ to give a differential polynomial rather than a rational function). This gives us (reordering the equations by rank): $x_1 - y, \dot{x}_1  - x_2, \dot{x}_2 - x_3, \cdots \dot{x}_{n-1} - x_n, \Psi(x_1, \dots, x_n)$
, where $\Psi$ is now written in terms of the $x_i$ rather than in terms of $y$.

Then simply following the reduction algorithm in \cite{Audoly2001} shows that the Ritt's pseudodivision method will generate a characteristic set which is just $\Psi$. This can be seen by noting that the leaders of the equations are $x_1$ for the measurement equation $y = x_1$, and the derivative terms on the left side for all the state equations. Then simply following the procedure of \cite{Audoly2001} will result in each reduction step replacing the $\dot{x}_1, \dots, \dot{x}_{n-1}$ with $\dot{y}, \dots, y^{(n-1)}$ in sequence, until we reach the $\dot{x}_n$ equation (i.e. $\Psi$). At this point, the system will be in the form:
$x_1 - y, \dot{y}  - x_2, \ddot{y} - x_3, \cdots,  y^{(n-1)} - x_n, \Psi(x_1, \dots, x_n)$.
Reduction of $\Psi$ with respect to each of the previous equations will simply substitute the appropriate number of derivatives of $y$ in for each $x_i$ as appropriate, returning our original equation $\Psi$ as the input-output equation for the system.

Now let us consider the multi-output case, $n_y = n > 1$. Let $\mb{\Psi} = \{\Psi_1, \dots, \Psi_n\}$ be our set of generalized input-output equations, assumed to be autoreduced. Order the $\Psi_i$ according their rank, and also order the output variables $y_1, \dots, y_n$ by the highest derivative of each that appears in $\mb \Psi$. We will denote the highest order (number of derivatives) of each output variable that is present in $\mb \Psi$ as $m_1 \leq \cdots \leq m_n$. Consider an arbitrary $\Psi_i$, and suppose its leader is some derivative (potentially of order 0) of $y_k$. The leader of $\Psi_i$ must be $y_k^{(m_k)}$, as otherwise some other $\Psi_j$ contains $y_k^{(m_k)}$ and $\Psi_j$ will not be reduced with respect to $\Psi_i$. Then since the $\Psi_i$ and $y_i$ are arranged in order of rank, we have that the leader of each $\Psi_i$ is simply $y_i^{(m_i)}$. Moreover, we note that no other $\Psi_j$ for $j\neq i$ can contain $y_i^{(m_i)}$, as then the two equations would not be reduced with respect to each other.

Now we can convert our system $\bf \Psi$ to first order by defining:
\begin{equation}
\begin{matrix} 
x_{11} = y_1,		& \cdots	& x_{1n} = y_n, \\
x_{21} = \dot{y}_1, 	& \cdots  	& x_{2n} = \dot{y}_n,\\
\vdots 			& 	& \vdots\\
x_{m_1 1} = y_1^{(m_1-1)}, &  & x_{m_n n} = y_n^{(m_n-1),}
\end{matrix}
\end{equation}
where we note that each column will be of equal length or longer as we move to the right (since $m_1 \leq \cdots \leq m_n$). We rank the $x_{ij}$'s lexicographically as: $x_{11} < x_{12} < \cdots < x_{1n} < x_{21}<  x_{22} < \cdots<  x_{2n} < \cdots < x_{m_1 1} < \cdots < x_{m_1 n} < \cdots < x_{m_n n}$. This allows us to rewrite our system as a first order set of differential polynomials (all equal to zero) in increasing rank order as:
\begin{equation}\label{eq:blob}
\begin{matrix} 
x_{11} - y_1,		& \cdots			& x_{1i} - y_i, 			&\cdots		& x_{1n} - y_n, \\
\dot{x}_{11} - x_{21}, & \cdots	  		& \dot{x}_{1n} - x_{2n}, 	&\cdots 		&\dot{x}_{1n} - x_{2n}, \\
\vdots 			& 				& \vdots				&			& \vdots\\
\dot{x}_{m_1-1,1} - x_{m_11}, 	& \cdots  	& \dot{x}_{m_1-1, i} - x_{m_1 i}, & \cdots  	& \dot{x}_{m_1-1, n} - x_{m_1 n}, \\
\Psi_1(\dot{x}_{m_1 1}), &  \cdots 		& \dot{x}_{m_1 i} - x_{m_1+1,i}, &  \cdots 	& \dot{x}_{m_1 n} - x_{m_1+1,n}, \\
					&			&\vdots				&	\cdots	&\vdots\\
					&			&\Psi_i(\dot{x}_{m_i i}),	&			&\vdots\\
					&			&					&			&\Psi_n(\dot{x}_{m_n n})
\end{matrix}
\end{equation}
where the $\Psi_i$ are rewritten in terms of the $x_{jk}$, with each $\Psi_i$ written as $\Psi_i(\dot{x}_{m_i i})$ to emphasize the highest rank variable (leader). We note that this system is in state-space form, since each $\Psi_i$ is the only equation to contain $\dot{x}_{m_i i} = y_i^{(m_i)}$ (the highest derivative of $y_i$). The equations in Eq.~\eqref{eq:blob} are ordered so that rank increases left to right and down the rows. 

We note that the equations in each column of Eq.~\eqref{eq:blob} up until each $\Psi_i$ only contain the leader (or any variables) from the equation immediately above them. This means that the reduction algorithm of \cite{Audoly2001} will proceed down each column independently, just as it would in the single input-output equation case, until $\Psi_1$ is reached. 

At this point, all previous differential polynomials in the list will be of the form $y_j^{(i-1)} - x_{ij}$, for $i\leq m_1$. We note that $\Psi_1(\dot{x}_{m_1 1})$ cannot depend on any subsequent $x_{ij}$ (with $j>m_1$) as this would mean the original $\Psi$ had a different leader. Then the reduction process will simply replace all the $x_{ij}$ with $y_j^{(i-1)}$, just as it would in the single output case (except now potentially using equations from more than one column). This will result in $\Psi_1$ being rewritten in it's original form, leaving the first column fully autoreduced with respect to itself. The process will continue until $\Psi_2$ is reached, at which point we note that the same procedure will occur. We note that because $\Psi_2$ cannot contain the leader of $\Psi_1$, $\dot{x}_{m_1 1} = y_1^{(m_1)}$, $\Psi_2$ will be reduced with respect to $\Psi_1$, and so this process will not alter $\Psi_1$ in any way. This will continue until the full system is autoreduced, resulting in:
\begin{equation}\label{eq:blob}
\begin{matrix} 
x_{11} - y_1,		& \cdots			& x_{1i} - y_i, 			&\cdots		& x_{1n} - y_n, \\
\dot{y}_{1} - x_{21}, & \cdots	  		& \dot{x}_{1n} - x_{2n}, 	&\cdots 		&\dot{x}_{1n} - x_{2n}, \\
\vdots 			& 				& \vdots				&			& \vdots\\
\dot{y}_1^{(m_1-1)} - x_{m_11}, 	& \cdots  	& \dot{y}_i^{(m_1-1)} - x_{m_1 i}, & \cdots  	& \dot{y}_n{(m_1-1)} - x_{m_1 n}, \\
\Psi_1, &  \cdots 		& \dot{y}_i^{(m_1)} - x_{m_1+1,i}, &  \cdots 	& \dot{y}_n^{(m_1)} - x_{m_1+1,n}, \\
					&			&\vdots				&	\cdots	&\vdots\\
					&			&\Psi_i,	&			&\vdots\\
					&			&					&			&\Psi_n
\end{matrix}
\end{equation}
namely, the initial transformation equations we started with, along with the input-output equations $\bf \Psi$, now in terms of only the observed variables $\bf y$, making the input-output equations of the characteristic set given by $\bf \Psi$ as desired.
\end{proof}

\pagebreak
\bibliographystyle{unsrt}
\bibliography{EisenbergRefs.bib}

\begin{thebibliography}{10}

\bibitem{Bellu2007}
G.~Bellu, M.~P. Saccomani, S.~Audoly, and L.~D'Angio.
\newblock Daisy: a new software tool to test global identifiability of
  biological and physiological systems.
\newblock {\em Comput Methods Programs Biomed}, 88(1):52--61, 2007.

\bibitem{Chappell1998}
Michael~J. Chappell and Roger~N. Gunn.
\newblock A procedure for generating locally identifiable reparameterisations
  of unidentifiable non-linear systems by the similarity transformation
  approach.
\newblock {\em Mathematical Biosciences}, 148(1):21--41, 1998.

\bibitem{Cobelli1980}
C.~Cobelli and J.~J. DiStefano.
\newblock Parameter and structural identifiability concepts and ambiguities: a
  critical review and analysis.
\newblock {\em American Journal of Physiology - Regulatory, Integrative and
  Comparative Physiology}, 239(1):R7--R24, 1980.

\bibitem{Pohjanpalo1978}
H.~Pohjanpalo.
\newblock System identifiability based on the power series expansion of the
  solution.
\newblock {\em Mathematical Biosciences}, 41(1-2):21--33, 1978.

\bibitem{Saccomani2003}
Maria Pia~Saccomani, Stefania Audoly, and Leontina D'AngiÚ.
\newblock Parameter identifiability of nonlinear systems: the role of initial
  conditions.
\newblock {\em Automatica}, 39(4):619--632, 2003.

\bibitem{Bellman1970}
R.~Bellman and K.~J. Astrom.
\newblock On structural identifiability.
\newblock {\em Mathematical Biosciences}, 7(3-4):329--339, 1970.

\bibitem{DiStefano1983}
J.~J. Distefano.
\newblock Complete parameter bounds and quasiidentifiability conditions for a
  class of unidentifiable linear systems.
\newblock {\em Mathematical Biosciences}, 65(1):51--68, 1983.

\bibitem{Ljung1994}
L.~Ljung and T.~Glad.
\newblock On global identifiability for arbitrary model parameterization.
\newblock {\em Automatica}, 30(2):265--276, 1994.

\bibitem{Meshkat2009}
N.~Meshkat, M.~Eisenberg, and J.~J. Distefano.
\newblock An algorithm for finding globally identifiable parameter combinations
  of nonlinear ode models using {Groebner} bases.
\newblock {\em Math Biosci}, 222(2):61--72, 2009.

\bibitem{Ollivier1990}
F.~Ollivier.
\newblock Le problem de l'identifiabilite structurelle globale: etude
  theoretique methodes effectives et bornes de complexite.
\newblock {\em PhD Thesis, Ecole Polytechnique}, 1990.

\bibitem{Audoly2001}
S.~Audoly, G.~Bellu, L.~D'Angio, M.~P. Saccomani, and C.~Cobelli.
\newblock Global identifiability of nonlinear models of biological systems.
\newblock {\em IEEE Trans Biomed Eng}, 48(1):55--65, 2001.

\bibitem{Ritt1950}
J.~F. Ritt.
\newblock {\em Differential Algebra}.
\newblock New York: Academic Press, 1950.

\bibitem{Margaria2001}
Gabriella Margaria, Eva Riccomagno, Michael~J. Chappell, and Henry~P. Wynn.
\newblock Differential algebra methods for the study of the structural
  identifiability of rational function state-space models in the biosciences.
\newblock {\em Mathematical Biosciences}, 174(1):1 -- 26, 2001.

\bibitem{hong2018global}
Hoon Hong, Alexey Ovchinnikov, Gleb Pogudin, and Chee Yap.
\newblock Global identifiability of differential models.
\newblock {\em arXiv preprint arXiv:1801.08112}, 2018.

\bibitem{Meshkat2012}
N.~{Meshkat}, C.~{Anderson}, and J.~J. {DiStefano}, III.
\newblock {Alternative to Ritt's Pseudodivision for finding the input-output
  equations in algebraic structural identifiability analysis}.
\newblock {\em ArXiv e-prints}, March 2012.

\bibitem{Mansfield1991}
Elizabeth Mansfield.
\newblock {\em Differential Gr\"{o}bner Bases}.
\newblock PhD thesis, University of Sydney, 1991.

\bibitem{meshkat2014finding}
Nicolette Meshkat, Christine Er-zhen Kuo, and Joseph DiStefano~III.
\newblock On finding and using identifiable parameter combinations in nonlinear
  dynamic systems biology models and combos: a novel web implementation.
\newblock {\em PLoS One}, 9(10):e110261, 2014.

\bibitem{Jacquez1985}
John~A. Jacquez and Peter Greif.
\newblock Numerical parameter identifiability and estimability: Integrating
  identifiability, estimability, and optimal sampling design.
\newblock {\em Mathematical Biosciences}, 77(1-2):201--227, 1985.

\bibitem{raksanyi1985identifiability}
Atilla Raksanyi, Yves Lecourtier, Eric Walter, and Alain Venot.
\newblock Identifiability and distinguishability testing via computer algebra.
\newblock {\em Mathematical biosciences}, 77(1-2):245--266, 1985.

\bibitem{walter1996identifiability}
Eric Walter and Luc Pronzato.
\newblock On the identifiability and distinguishability of nonlinear parametric
  models.
\newblock {\em Mathematics and computers in simulation}, 42(2-3):125--134,
  1996.

\bibitem{distefano2015dynamic}
Joseph DiStefano~III.
\newblock {\em Dynamic systems biology modeling and simulation}.
\newblock Academic Press, 2015.

\bibitem{eisenberg2013identifiability}
Marisa~C Eisenberg, Suzanne~L Robertson, and Joseph~H Tien.
\newblock Identifiability and estimation of multiple transmission pathways in
  cholera and waterborne disease.
\newblock {\em Journal of theoretical biology}, 324:84--102, 2013.

\bibitem{walch2016parameter}
Olivia~J Walch and Marisa~C Eisenberg.
\newblock Parameter identifiability and identifiable combinations in
  generalized hodgkin--huxley models.
\newblock {\em Neurocomputing}, 199:137--143, 2016.

\bibitem{kao2018practical}
Yu-Han Kao and Marisa~C Eisenberg.
\newblock Practical unidentifiability of a simple vector-borne disease model:
  Implications for parameter estimation and intervention assessment.
\newblock {\em Epidemics}, 25:89--100, 2018.

\bibitem{meshkat2014identifiable}
Nicolette Meshkat and Seth Sullivant.
\newblock Identifiable reparametrizations of linear compartment models.
\newblock {\em Journal of Symbolic Computation}, 63:46--67, 2014.

\bibitem{vajda1989similarity}
Sandor Vajda, Keith~R Godfrey, and Herschel Rabitz.
\newblock Similarity transformation approach to identifiability analysis of
  nonlinear compartmental models.
\newblock {\em Mathematical biosciences}, 93(2):217--248, 1989.

\bibitem{Cox1996}
D.~O'Shea D.~Cox, J.~Little.
\newblock {\em Ideals, Varieties, and Algorithms: An Introduction to
  Computational Algebraic Geometry and Commutative Algebra}.
\newblock Springer, 1996.

\bibitem{Aistleitner2010}
Christian Aistleitner.
\newblock Relations between gr\"{o}bner bases, differential gr\"{o}bner bases,
  and differential characteristic sets.
\newblock {\em Thesis, Institut f\"{u}r Symbolisches Rechnen}, 2010.

\bibitem{Faugere1999}
Jean-Charles Faugere.
\newblock A new efficient algorithm for computing groebner bases (f4).
\newblock {\em Journal of Pure and Applied Algebra}, 139(1-3):61 -- 88, 1999.

\bibitem{Collart1993}
S.~Collart, M.~Kalkbrener, and D.~Mall.
\newblock The gršbner walk.
\newblock Technical report, J. Symbolic Computation, 1993.

\bibitem{CarraFerro1987}
G.~Carra'Ferro.
\newblock Groebner bases and differential algebra.
\newblock In {\em Proceedings of the 5th international conference, AAECC-5 on
  Applied Algebra, Algebraic Algorithms and Error-Correcting Codes}, pages
  129--140, New York, NY, USA, 1987. Springer-Verlag New York, Inc.

\end{thebibliography}
\end{document}